%% file: XrayHaloProfile_ApJ.tex
\documentclass[twocolumn]{aastex61}

\newcommand{\nh}{$N_\mathrm{H}$}
\newcommand{\nhI}{$N_\mathrm{HI}$}

\newcommand{\nhtwo}{$N_\mathrm{H_{2}}$}
\newcommand{\Tab}[1]{Table \ref{#1}}
\newcommand{\Fig}[1]{Figure \ref{#1}}
\newcommand{\Sec}[1]{Section \ref{#1}}

\newcommand{\kThalo}{$kT_\mathrm{halo}$}
\newcommand{\EMhalo}{$EM_\mathrm{halo}$}
\newcommand{\Fehalo}{[O/Fe]$_\mathrm{halo}$}
\newcommand{\EMlocal}{$EM_\mathrm{local}$}
\newcommand{\Ncxb}{$N_\mathrm{CXB}$}
\newcommand{\EMunit}{~cm$^{-6}$~pc}

\received{}
\revised{}
\accepted{}
\submitjournal{ApJ}

\shorttitle{}
\shortauthors{Nakashima et al.}

\begin{document}

\title{Spatial distribution of the Milky Way hot gaseous halo constrained by Suzaku X-ray observations}

\email{shinya.nakashima@riken.jp}

\author{Shinya Nakashima}
\affil{RIKEN High Energy Astrophysics Laboratory, 2-1 Hirosawa, Wako, Saitama, 351-0198, Japan}

\author{Yoshiyuki Inoue}
\affil{RIKEN Interdisciplinary Theoretical and Mathematical Sciences (iTHEMS) Program, 2-1 Hirosawa, Wako, Saitama, 351-0198, Japan}

\author{Noriko Yamasaki}
\affil{Institute of Space and Astronautical Science, Japan Aerospace Exploration Agency, 3-1-1 Yoshinodai, Chuo-ku, Sagamihara, Kanagawa 252-5210, Japan}

\author{Yoshiaki Sofue}
\affil{Institute of Astronomy, The University of Tokyo, Mitaka, Tokyo 181-0015, Japan}

\author{Jun Kataoka}
\affil{Research Institute for Science and Engineering, Waseda University, 3-4-1 Okubo, Shinjuku, Tokyo 169-8555, Japan}

\author{Kazuhiro Sakai}
\affil{NASA Goddard Space Flight Center, Code 662, Greenbelt, MD20771, USA}



\begin{abstract}
The formation mechanism of the hot gaseous halo associated with the Milky Way Galaxy is still under debate. 
We report new observational constraints on the gaseous halo using 107 lines-of-sight of the Suzaku X-ray observations at $75\arcdeg<l<285\arcdeg$ and $|b|>15\arcdeg$ with a total exposure of 6.4~Ms.
The gaseous halo spectra are represented by a single-temperature plasma model in collisional ionization equilibrium.
The median temperature of the observed fields is 0.26~keV ($3.0\times10^6$~K) with a typical  fluctuation of $\sim30$\%.
The emission measure varies by an order of magnitude and marginally correlates with the Galactic latitude.
Despite the large scatter of the data, the emission measure distribution is roughly reproduced by a disk-like density distribution with a scale length of $\sim7$~kpc, a scale height of $\sim2$~kpc, and a total mass of $\sim5\times10^7$~$M_{\sun}$.
In addition, we found that a spherical hot gas with the $\beta$-model profile hardly contributes to the observed X-rays but that its total mass might reach $\gtrsim10^9$~$M_{\sun}$.
Combined with indirect evidence of an extended gaseous halo from other observations, the hot gaseous halo likely consists of a dense disk-like component and a rarefied spherical component; the X-ray emissions primarily come from the former but the mass is dominated by the latter. 
The disk-like component likely originates from stellar feedback in the Galactic disk due to the low scale height and the large scatter of the emission measures.
The median [O/Fe] of $\sim0.25$ shows the contribution of the core-collapse supernovae and supports the stellar feedback origin.
\end{abstract}





\section{Introduction} \label{sec:intro}
The evolution of galaxies is regulated by inflowing gas from the intergalactic medium and outflowing gas from the disk region \citep[][and references therein]{2017ARA&amp;A..55..389T}.
In a spiral galaxy with a mass of $\gtrsim10^{12}$~M$_\sun$, inflowing gas is expected to form a hot gaseous halo ($T > 10^6$~K) via accretion shocks and adiabatic compression 
extending to the viral radius \citep{2009MNRAS.395..160K,2010MNRAS.407.1403C,2012ApJ...759..137J}, while stellar feedback forms superbubbles in the disk and drives multi-phase gas outflows up to several kpc above the disk \citep{Hill_2012,2018ApJ...853..173K}.
Numerical simulations show divergent behavior in the formation of gaseous halos due to different implementations of feedback and star formation \citep{2017ApJ...843...47S}.
Therefore, observational constraints on the properties of hot gaseous halos are essential to understanding the amount of accreting and outflowing gas.
 
A hot gaseous halo around the Milky Way Galaxy (hereafter MW) has been confirmed via X-ray observations.
Early X-ray missions found a diffuse X-ray background in the 0.5--1.0~keV band \citep[and references therein]{1977SSRv...20..815T,1990ARA&amp;A..28..657M}, and the ROSAT all-sky survey revealed the detailed spatial distribution of the X-ray emissions \citep{1997ApJ...485..125S};
in addition to the prominent features around the center of the MW, significant excesses that cannot be explained by the superposition of extragalactic active galactic nuclei are found. 
After the advent of grating spectrometers and microcalorimeters, absorption and emission lines of \ion{O}{7} and \ion{O}{8} at zero-redshift were observed, which provide evidence of the association of hot gas with the MW \citep[e.g.,][]{2002ApJ...572L.127F,2002ApJ...576..188M}. 

The detailed spatial distribution of the MW hot gaseous halo has been extensively investigated using emission and absorption lines over the past decade.
Combined with emission and absorption line measurements toward LMC X-3, \cite{2009ApJ...690..143Y} constructed a disk-like distribution model with a scale height of a few kpc, suggesting a significant contribution of hot gas produced by stellar feedback rather than accretion shocks.
Similar results were obtained in other two lines-of-sight \citep{2010PASJ...62..723H,2014PASJ...66...83S}.
Conversely, \cite{2012ApJ...756L...8G} presented a hot gas distribution extending up to $\sim100$~kpc using absorption lines data toward several extragalactic sources. 
\cite{2013ApJ...770..118M,2015ApJ...800...14M} analyzed 29 absorption-lines and 649 emission-lines measurements and formulated an extended spherical morphology represented by the $\beta$ model.
The cause of the discrepancy between these results is not clear but might be due to different assumptions, such as the temperature profile and metallicity.

As a complement to the above line data analyses, broadband X-ray spectroscopy including the continuum has been performed.
In contrast to the line data analyses, where temperature and metallicity need to be assumed, 
broadband spectroscopy can self-consistently determine the temperature, emission measure, and metallicity.
This approach has been widely used for observations of nearby dark clouds with CCD detectors \citep{2007PASJ...59S.141S,2007ApJ...658.1081G,2007ApJ...661..304H,2015ApJ...799..117H}. 
However, systematic analyses with large samples are limited due to limited photon statistics compared to line measurements. 
\cite{2009PASJ...61..805Y} analyzed 13 lines-of-sight of the Suzaku observations, and \cite{2013ApJ...773...92H} analyzed  110 lines-of-sight of the XMM-Newton observations.

In this paper, we present new broadband spectroscopic results of the MW hot gaseous halo using 107 lines-of-sight of the Suzaku observations. 
The X-ray CCDs aboard Suzaku \citep[XIS;][]{2007PASJ...59S..23K} have low and stable instrumental background and  good spectral responses below 1~keV compared to the X-ray CCDs aboard Chandra and XMM-Newton \citep{2007PASJ...59S...1M}.
Therefore, it is suitable for the spectroscopy of faint diffuse emission.
The data selection and screening are explained in \Sec{sec:obs}.
The spectral modeling and results are shown in \Sec{sec:ana}, and the
interpretations of the results are discussed in \Sec{sec:discus}.   

\section{Observations and data reduction} \label{sec:obs}
\subsection{Data selection from the Suzaku archive} 

\begin{figure}
\epsscale{1.2}
\plotone{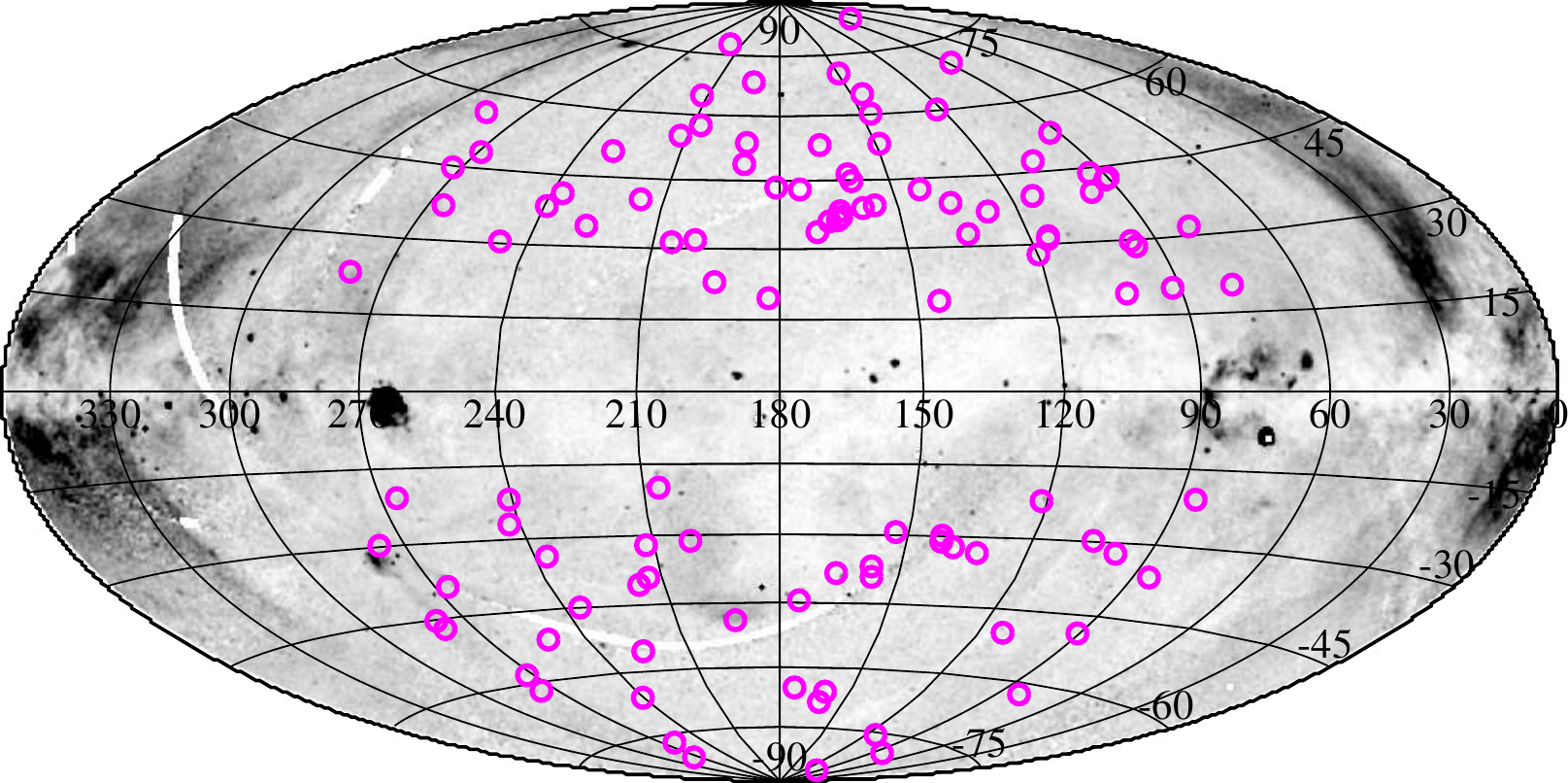}
\caption{Directions of the 107 fields analyzed in this paper overlaid on a gray scale image of  the ROSAT R45 band.
The fields are located at $75\arcdeg < l < 285\arcdeg$ and $|b| > 15\arcdeg$.
The size of the circles is artificial, and the actual Suzaku FoV is $17\farcm8 \times 17\farcm8$.
\label{fig:pointings}}
\end{figure}

We used  archival data of the Suzaku/XIS, which is sensitive to 0.2--12.0~keV X-rays. 
The XIS consists of three front-illuminated (FI) type CCDs (XIS0, XIS2, and XIS3) and one back-illuminated (BI) type CCD (XIS1) located at the focal planes of four independent X-ray telescopes \citep{2007PASJ...59S...9S}.
XIS2 has not been functioning since 2006 November, and was not used in our analysis.
The effective area at 1.5~keV is $\sim1030$~cm$^2$ combined with the remaining three sensors.
The field-of-view (FoV) is $\sim18\arcmin\times18\arcmin$ with a spatial resolution of $\sim2\arcmin$ in a half-power diameter.

We accumulated the observations pointing to the Galactic anticenter ($75\arcdeg < l < 285\arcdeg$) and outside the Galactic plane ($|b| > 15\arcdeg$).
Observations toward the Galactic center were not used because additonal diffuse hot gas associated with past Galactic Center activities would contaminate the results for the hot gaseous halo \citep[e.g.,][]{2010ApJ...724.1044S,2013ApJ...773...20N,2013ApJ...779...57K,2016ApJ...829....9M}.

The observations contaminated by other X-ray emitting extended objects such as clusters of galaxies, galaxies,  supernova remnants, and superbubbles were excluded.
Bright compact objects are other contaminant sources due to the wide point spread functions of the Suzaku telescopes.
Referring to the HEASARC Master X-ray Catalog\footnote{\url{https://heasarc.gsfc.nasa.gov/W3Browse/all/xray.html}}, we removed observations aimed at sources brighter than $10^{11}$~erg~s~cm$^{-2}$. 
Photons from extremely bright sources outside the XIS FoV are also scattered into the detector; that is so-called "stray light" \citep{2007PASJ...59S...9S}.
Therefore, observations within a $90\arcmin$ radius of sources of $F_\mathrm{X} > 10^{10}$~erg~s~cm$^{-2}$ were discarded.
In addition, the observations of specific targets, such as the Moon, Jupiter, nearby dark clouds (MBM16, MBM20, and LDN1563), and helium focusing cone were excluded.
Finally, observations of which effective exposures were less then 10~ks after the screening described in the next section were also excluded.
As a result of these selections, we accumulated 122 observations with a total exposure of $\sim6.4$~Ms (\Fig{fig:pointings} and \Tab{tab:obsresults}).
Some observations cover the same sky regions.
To identify line-of-sight directions, we assigned the region IDs; the same region ID was assigned to the observations of the same sky region.
The total number of lines-of-sight is 107.

\subsection{Data reduction} 
The selected XIS data were reprocessed via the standard pipeline with HEASOFT version 6.22 and the calibration database as of 2016 April 1.
We then removed additional flickering pixels that were found in the long-term background monitering\footnote{\url{http://www.astro.isas.ac.jp/suzaku/analysis/xis/nxb_new2/}}.
Due to the charge leakage, segment A of the XIS0 was not used  for the data taken after 2009 June 27\footnote{\url{http://www.astro.isas.ac.jp/suzaku/analysis/xis/xis0_area_discriminaion/}}.

\begin{figure}[t]
\plotone{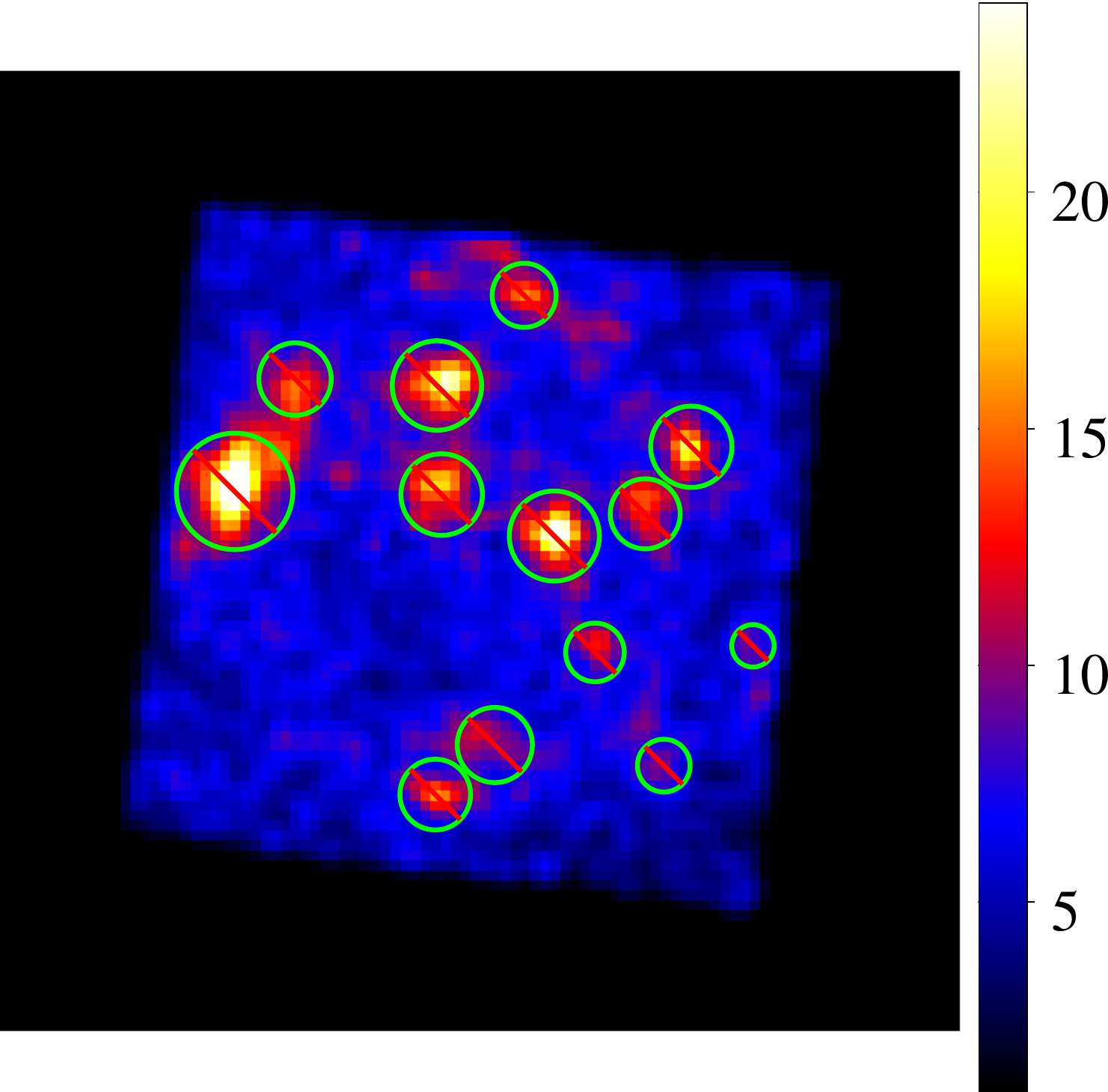}
\caption{XIS count image in the 0.7--5.0~keV band for OBSID 502047010 (region \#7) aiming at ($l$, $b$) = ($86\fdg0$, $-20\fdg8$).
The data of XIS\,0, XIS\,1, and XIS\,3 were co-added.
The vignetting effect was not corrected, and Gaussian smoothing with $\sigma = 16\arcsec$ was applied.
Point source candidates detected by the \texttt{wavdetect} tool are shown in the green circles.
\label{fig:wavdetect}}
\end{figure}

To remove point sources in the XIS FoVs, we created 0.7--5.0~keV raw count images where data from all the sensors were co-added and searched for source candidates using the {\tt wavdetect}  tool in the CIAO package\footnote{\url{http://cxc.harvard.edu/ciao/}}.
\Fig{fig:wavdetect} shows an example of the results.
We found 13 candidates in that image including possible false detections with a significance of $<3\sigma$.
The number of detected sources in one observation ranged between 1 and 16, depending on the effective exposure times.
All the detected candidates from the event list were removed via circular regions (\Fig{fig:wavdetect}).

The geocoronal solar wind charge exchange (SWCX) emission is a possible contaminant source below 1~keV.
Its flux varies on a time scale of hours and correlates with the proton flux of the solar wind \citep[e.g.,][]{2007PASJ...59S.133F,2013PASJ...65...63I}. 
Previous studies have successfully reduced geocoronal SWCX contamination by screening out  durations  where the solar wind proton flux exceeds $4\times10^8$~protons~cm$^{-2}$~s$^{-1}$ \citep[e.g.,][]{2009PASJ...61..805Y,2013ApJ...773...92H,2015ApJ...800...14M}.
We followed these screening criteria with the proton flux calculated from the OMNI database\footnote{\url{http://omniweb.gsfc.nasa.gov/ow.html}}.

For the data below 0.7~keV, we adopted one additional screening criterion to suppress contamination by the \ion{O}{1}~K$_{\alpha}$ emission from the sunlit Earth atmosphere. 
As reported by \cite{2014PASJ...66L...3S},  the \ion{O}{1}~K$_{\alpha}$ contamination has become significant since 2011 even after excluding periods where the angle between the satellite pointing direction and the sunlit Earth's rim ($\mathrm{DYE\_ELV}$) is less than $20\arcdeg$.
This phenomenon is likely due to increasing solar activity. 
The \ion{O}{1}~K$_{\alpha}$ contamination can be reduced by applying a higher $\mathrm{DYE\_ELV}$ threshold.
Therefore, we used a new $\mathrm{DYE\_ELV}$ threshold for individual observations;
we calculated the 0.5--0.6 keV count rates for each $\mathrm{DYE\_ELV}$ with a binning of $10\arcdeg$, and determined the $\mathrm{DYE\_ELV}$ threshold where the count rate significantly increases.
The actual values are listed in \Tab{tab:obsresults}.
The standard screening criteria ($\mathrm{DYE\_ELV} > 20\arcdeg$) is still applicable to some observations.

We extracted spectra from  0.4--0.7~keV and 0.7--5.0~keV separately;
the day-Earth screening was only applied to the former spectra.
Spectra of the FI CCDs (XIS0 and XIS3) were co-added to increase the photon statistics. 
The Instrumental backgrounds were estimated from the night-Earth database using {\tt xisnxbgen} \citep{2008PASJ...60S..11T}.
Energy responses of each sensor were generated by {\tt xisrmfgen} and {\tt xissimarfgen} \citep{2007PASJ...59S.113I}.


\input{table_pub_table_obs_results}

\section{Analysis and results} \label{sec:ana}
We constructed a spectral model  (\Sec{sec:spec-model}) and fitted it to the data to derive the parameters of the hot gaseous halo (\Sec{sec:spec-fit}).
Correlations between the parameters were also investigated (\Sec{sec:anal:correlation}).

\subsection{Spectral model}
\label{sec:spec-model}
Our spectral model consisted of three components: the hot gaseous halo, the local emission component, and the cosmic X-ray background (CXB).
This model is similar to those used in previous broadband spectroscopy of the soft X-ray background \citep[e.g.,][]{2013ApJ...773...92H} but it included recent updates of the atomic database, the solar metallicity, and the Galactic hydrogen column density as described below.

The hot gaseous halo component is described by a single temperature plasma in collisional ionization equilibrium (CIE).
We used the APEC plasma spectral model \citep{2012ApJ...756..128F} with AtomDB version 3.0.9.
The latest solar abundance table of \citet{2009LanB...4B...44L} was adopted as a reference of the metallicity. 
In the spectral fitting, the plasma temperature (\kThalo{}) and the emission measure (\EMhalo{}) were treated as free parameters.
The metallicity is difficult to determine in a CCD spectrum because lines and radiative recombination continua from oxygen and iron exceed bremsstrahlung from hydrogen in $kT=0.2$~keV plasma.
Therefore, we allowed only the iron abundance ($Z_\mathrm{Fe}$) to vary and fixed the other metal abundances to the solar values.
The setting allowed us to obtain the abundance ratio of oxygen to iron (\Fehalo{} $= \log_{10}(Z_\mathrm{O}/Z_\mathrm{Fe})$).
When $Z_\mathrm{Fe}$ was not constrained within 0.1--10 times the solar value during the fitting procedure, we fixed it to the solar value.
Previously, several studies have assumed a metallicity of 0.3 solar instead of the solar value for the hot gaseous halo \citep[e.g.,][]{2015ApJ...800...14M}.
We confirmed that fixing the abundances (except iron) to 0.3 solar increases \EMhalo{} by a factor of 3 without affecting the other parameters.


The local emission originates from the local hot bubble and the heliospheric SWCX \citep[e.g.,][]{2007PASJ...59S.133F,2017ApJ...834...33L}.
Despite their debatable physical properties, a spectrum is empirically described by a single CIE plasma of $kT\sim0.1$~keV with the solar metallically in the CCD spectra \citep[e.g.,][]{2007PASJ...59S.141S,2009PASJ...61..805Y,2013ApJ...773...92H}.  
We used the same phenomenological model; the temperature was fixed to 0.1~keV and the emission measure (\EMlocal{}) was allowed to vary.

The cosmic X-ray background (CXB) is a superposition of unresolved extragalactic sources.
An absorbed power-law function with a photon index of 1.45 represents the CXB spectrum in the 0.3--7~keV band \citep{2017ApJ...837...19C}. 
The normalization of the power-law function at 1~keV (\Ncxb{}) is $\sim$10~photons~cm$^{-2}$~s$^{-1}$~sr$^{-1}$~keV$^{-1}$, but spatially fluctuates by $\sim$15\% for the XIS FoV ($\sim0.8$~deg$^2$) due to the cosmic variance \citep{2009A&amp;A...493..501M}. 
Therefore, we treated \Ncxb{} as a free parameter in our spectral model.

The hot gaseous halo emission and the CXB are subject to absorption due to the Galactic cold interstellar medium.
This absorption was modeled using TBabs code version 2.3 \citep{2000ApJ...542..914W}, in which cross sections of dust grains and molecules are taken into account.
The absorption hydrogen column density (\nh{}) of each line-of-sight was fixed to the value estimated by \cite{2013MNRAS.431..394W}, in which the contribution of not only neutral hydrogen atoms (\nhI{}) but also molecular hydrogen (\nhtwo{}) were included.
We confirmed that using only the \nhI{} values from \cite{2005A&amp;A...440..775K}, which has been widely used in previous studies, has no significant impact on our results.

\begin{figure}
\epsscale{1.1}
\plotone{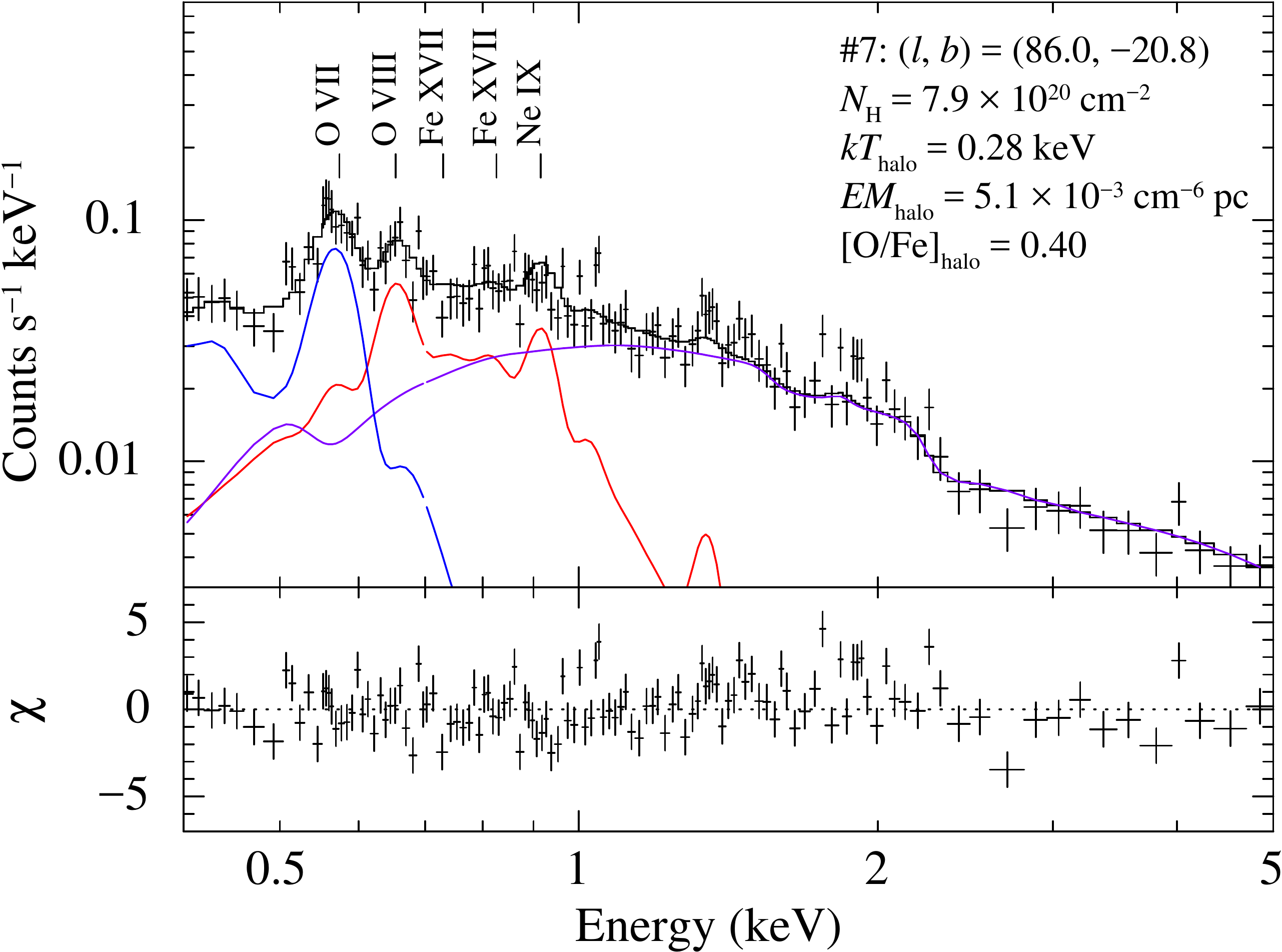}
\caption{The upper panel shows the XIS1 spectrum of OBSID 502047010 (region \#7).
For plotting purposes, the spectrum is binned so that each bin has at least 25 counts after subtracting the instrumental background.
The black curve is the best-fit model, which consists of three components: the Galactic hot gaseous halo (red), the local emission component (blue), and the CXB (purple).
The lower panel shows the residuals between the data and the model.
\label{fig:onespectrum}}
\end{figure}

\subsection{Spectral fitting results}
\label{sec:spec-fit}

Spectral fitting was performed with Xspec version 12.9.1n.
Spectra in the same region IDs were simultaneously fitted. 
The best-fit parameters were determined by minimizing the C-statistic \citep{1979ApJ...228..939C} with a Poisson background\footnote{referred to as the "W-statistics" in the Xspec manual (\url{https://heasarc.nasa.gov/xanadu/xspec/manual/XSappendixStatistics.html})}.
To compensate for the systematic differences in the effective areas among the sensors \citep{2011A&amp;A...525A..25T}, we allowed the relative normalization to vary between the BI and FI spectra.


\Fig{fig:onespectrum} shows an example of the fitting results.
The local emission (blue curve) dominates the spectrum below 0.6~keV whereas the CXB (purple curve) dominates the spectrum above 1.2~keV.
The hot gaseous halo emission (red curves) fills the remaining excess in the range of 0.6--1.0~keV.
The derived halo parameters, \kThalo{}$=0.28$~keV and \Fehalo{}$=0.40$, are primarily constrained by the emission lines of \ion{O}{7}, \ion{O}{8}, \ion{Fe}{17}, and \ion{Ne}{9}.
\Tab{tab:obsresults} summarizes the best-fit parameters for all the regions.



\begin{figure*}
\gridline{
	\leftfig{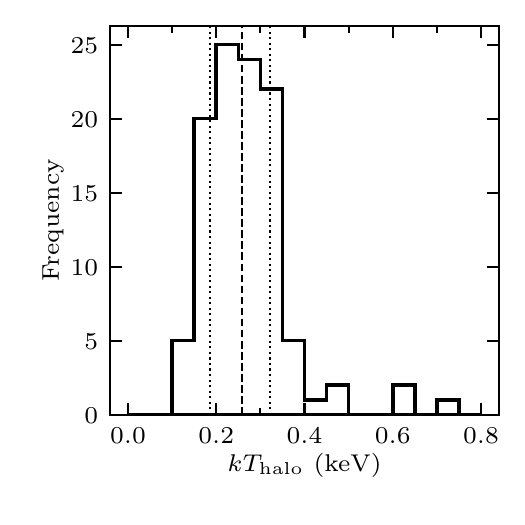}{0.32\textwidth}{}
	\fig{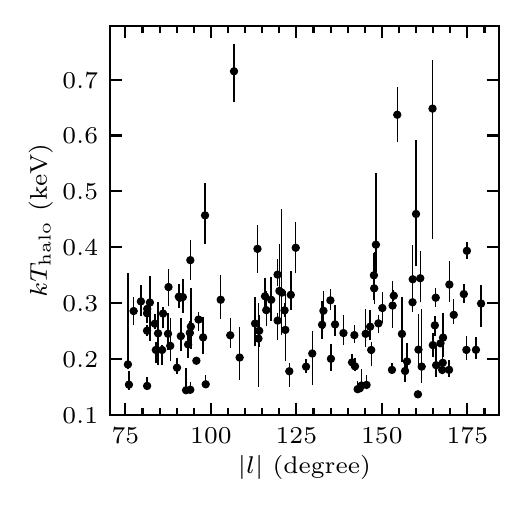}{0.32\textwidth}{}
	\rightfig{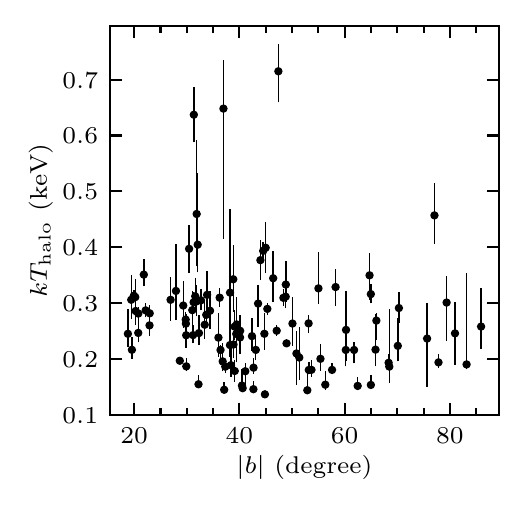}{0.32\textwidth}{}
}
\caption{(Left) Histogram of \kThalo{} derived from the spectral fitting.
The vertical dashed line indicates the median, and the vertical dotted lines indicate the 16th and 84th percentiles.
(Middle and Right) \kThalo{} versus $|l|$ and $|b|$, respectively.
\label{fig:kThalo}}
\end{figure*}

\begin{figure*}
\epsscale{1.1}
\gridline{
	\leftfig{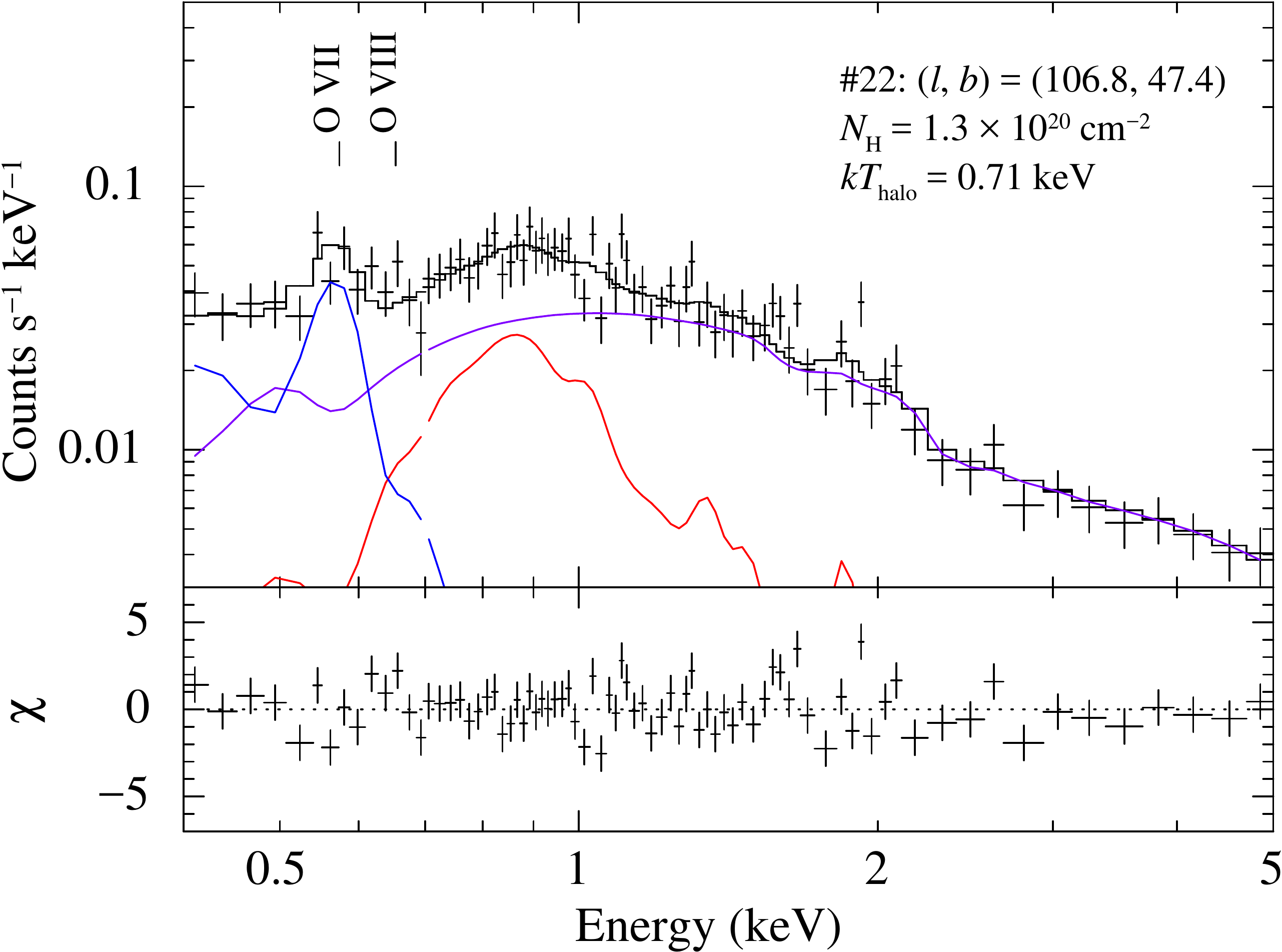}{0.33\textwidth}{}
	\fig{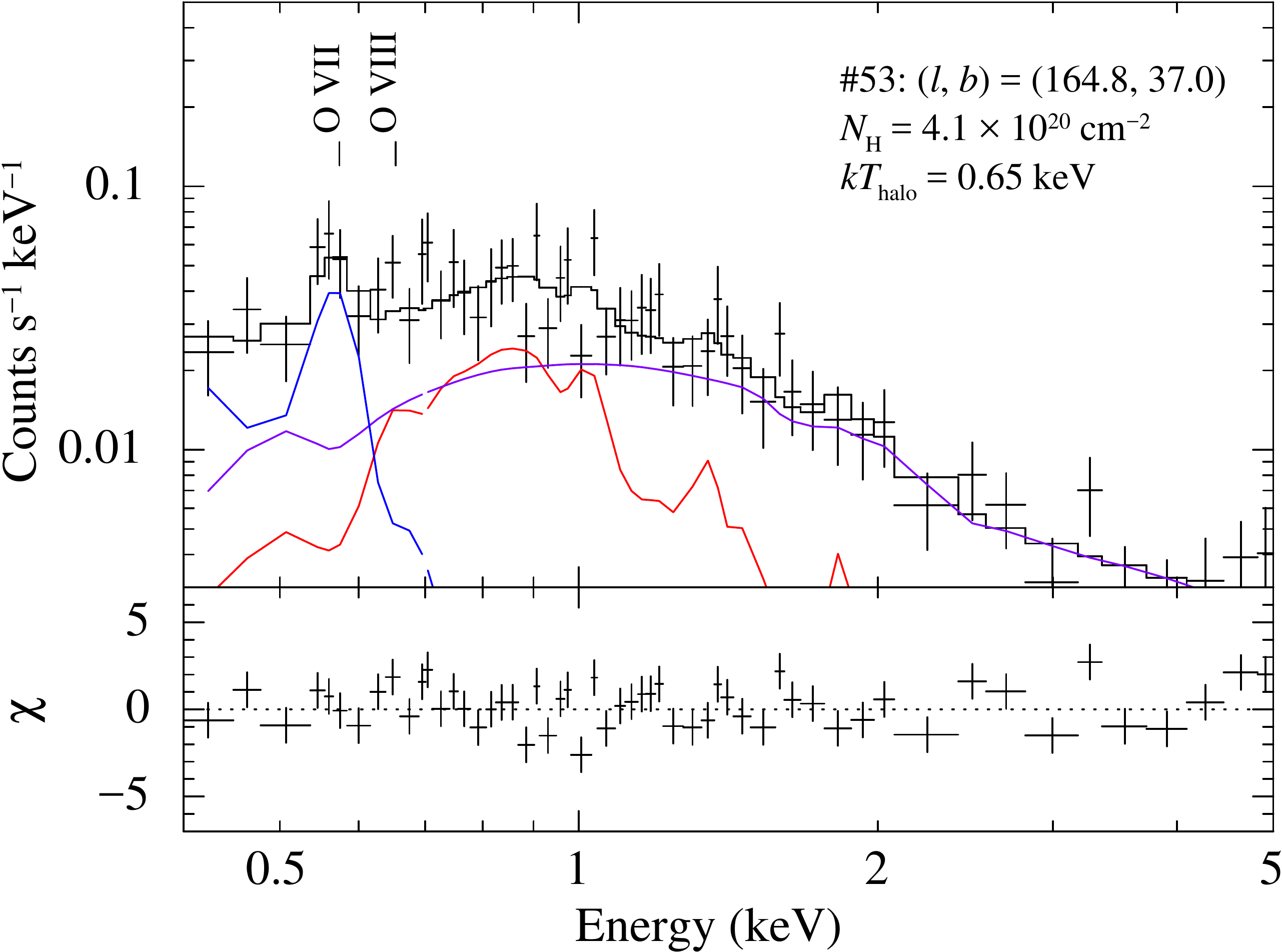}{0.33\textwidth}{}
	\rightfig{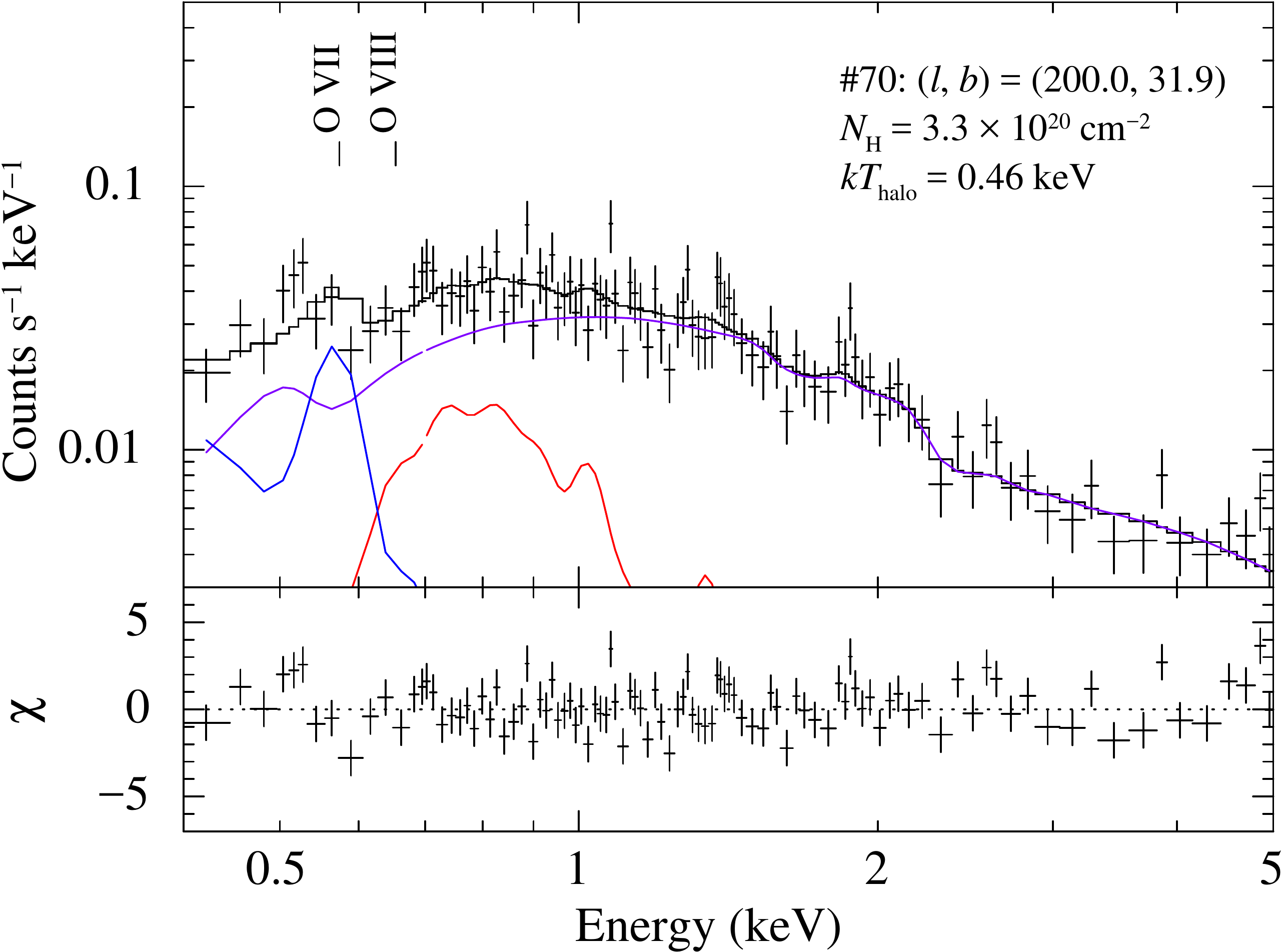}{0.33\textwidth}{}
}
\gridline{
	\leftfig{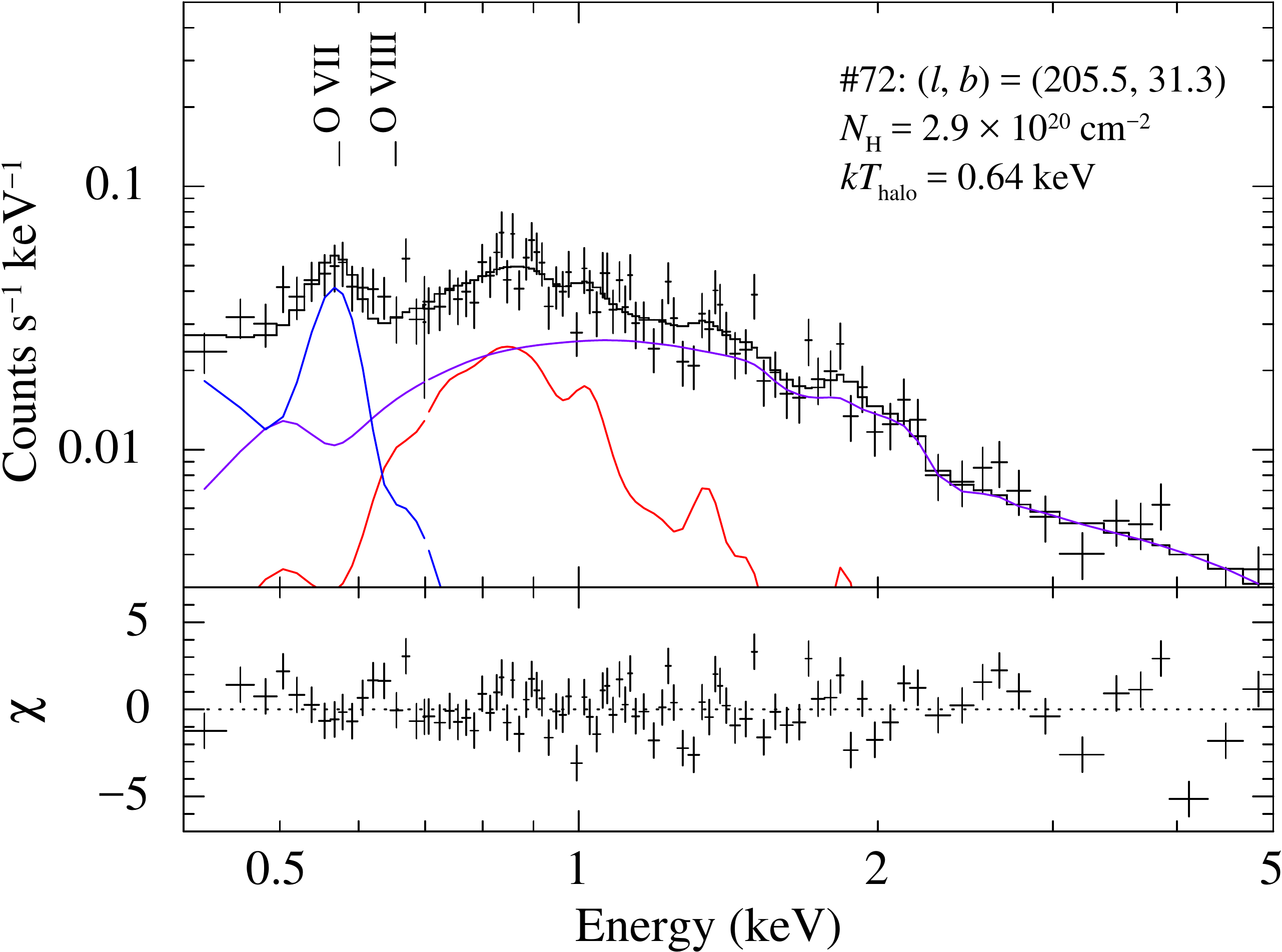}{0.33\textwidth}{}
	\fig{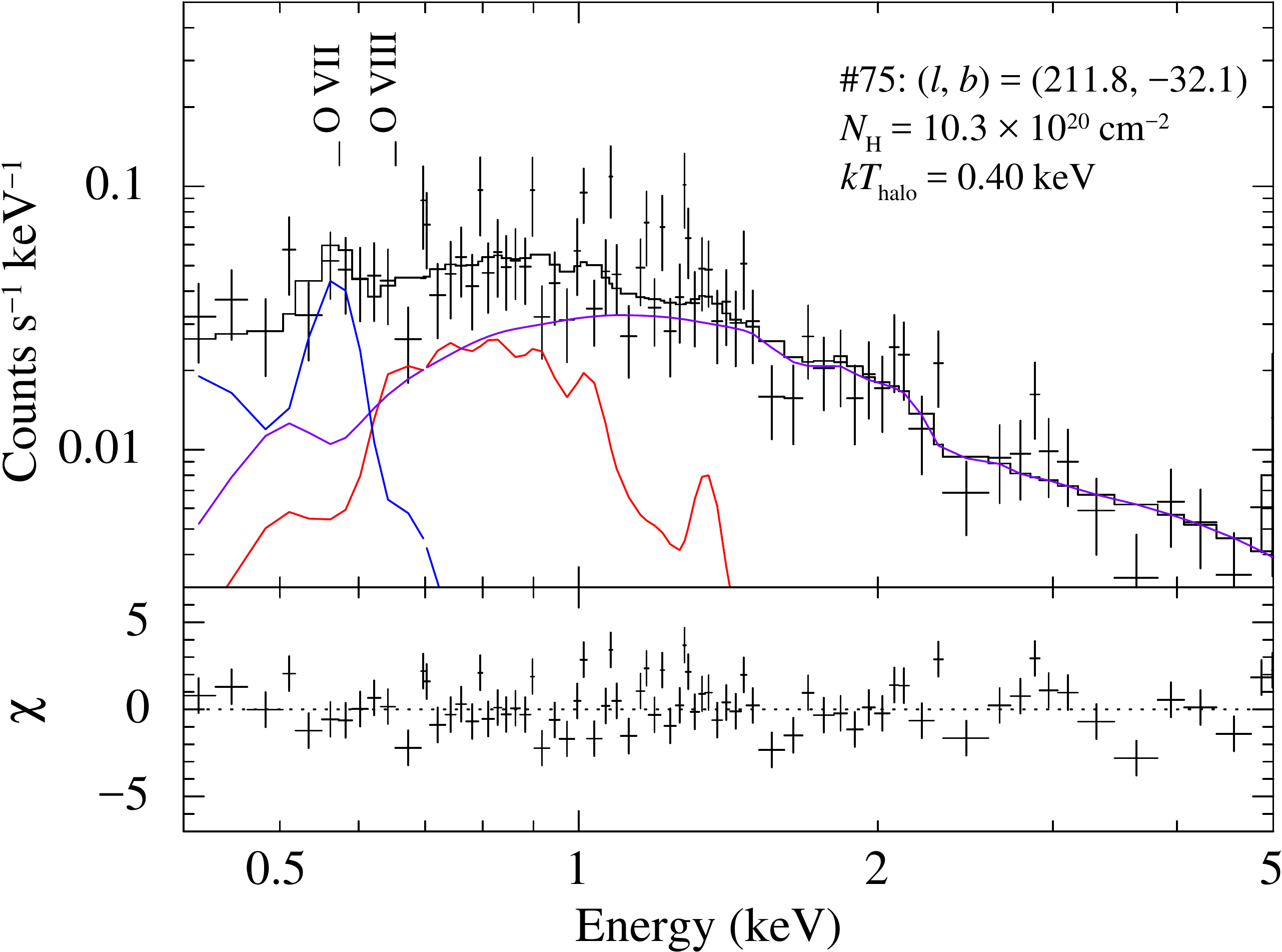}{0.33\textwidth}{}
	\rightfig{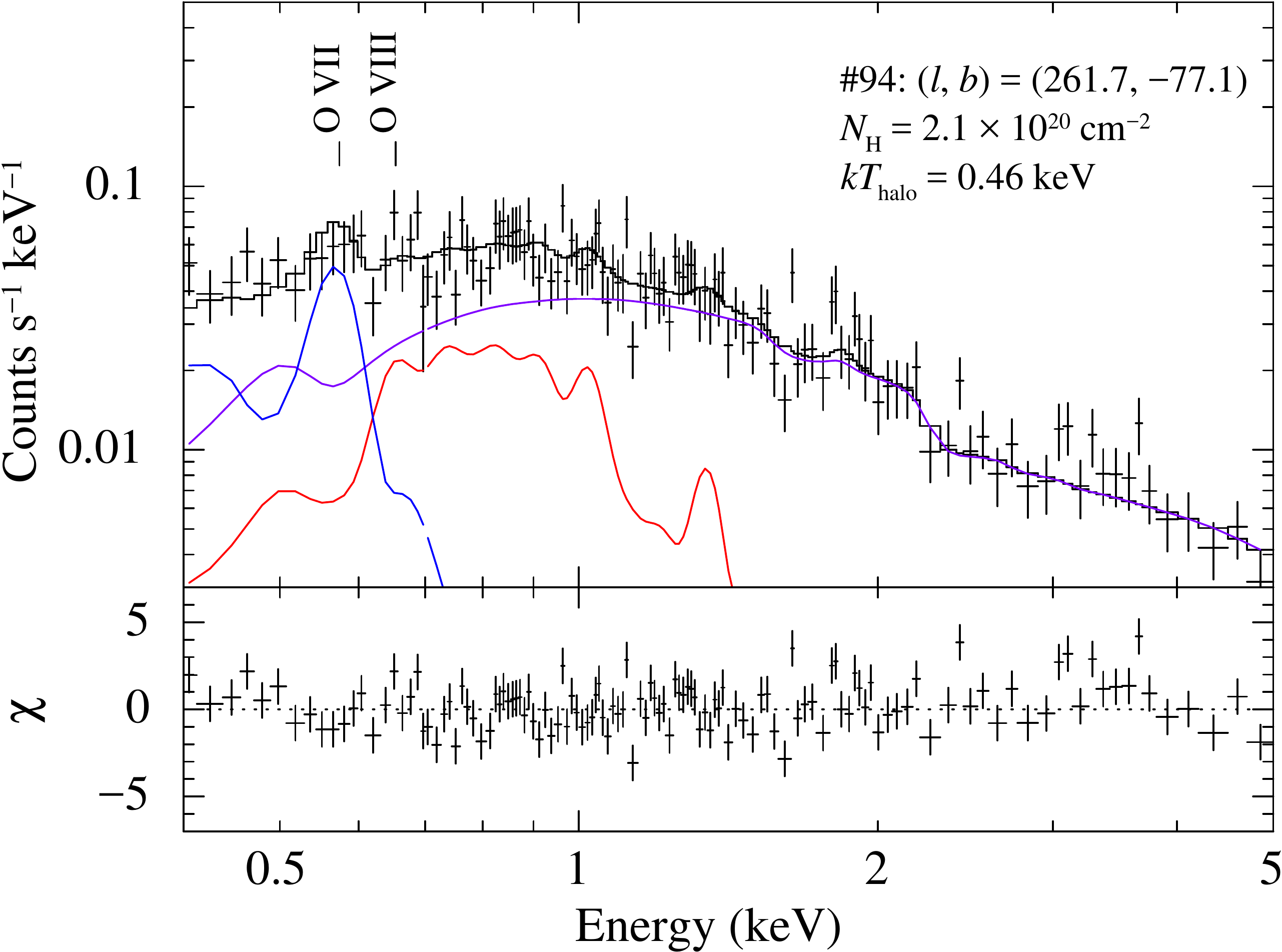}{0.33\textwidth}{}
}
\caption{Same as \Fig{fig:onespectrum} but for regions \#22, 53, 70, 72, 75, and 94, which shows \kThalo{} $>0.4$~keV.
\label{fig:highTspectra}}
\end{figure*}

A histogram of the best-fit \kThalo{} is shown on the left side of \Fig{fig:kThalo}.
The median is 0.26~keV and the 16--84th percentile range is 0.19--0.32~keV.
The shape of the distribution is nearly symmetric with respect to the median value; however, six regions show significantly high temperatures ($kT > 0.4$~keV).
Spectra of these high-temperature regions are shown in \Fig{fig:highTspectra}.
They exhibit an excess of Fe L-shell lines between 0.7--0.9~keV and no clear \ion{O}{8}~Ly$\alpha$ line.
That is because the best-fit temperatures of these regions are higher than those of other regions. 
The lack of an \ion{O}{8}~Ly$\alpha$ line is not caused by interstellar absorption because the transmission of \ion{O}{8}~Ly$\alpha$ is $>50$\% for those regions, where \nh{} is in the range of  1.3--$10.3\times10^{20}$~cm$^{-2}$.

The middle and the right side of \Fig{fig:kThalo} show \kThalo{} versus $|l|$ and $|b|$, respectively, where $|l|$ is defined as 
\begin{eqnarray}
|l|  & = &
\left\{
\begin{array}{ll}
l & (0\arcdeg \leq l < 180\arcdeg )\\
360\arcdeg - l \hphantom{aa}& (180\arcdeg \leq l < 360\arcdeg ).
\end{array}  \right. 
\end{eqnarray}
Spearman rank correlations for those two plots are shown in \Tab{tab:correlation}.
We found a marginal negative correlation of $\rho=-0.24$ between \kThalo{} and $|b|$ with a $p$-value of 0.019. 
Conversely, no correlation was observed between \kThalo{} and $|l|$.

\begin{deluxetable}{lcclcc}
\tablecaption{Spearman rank correlation between the halo parameters and the Galactic coordinates.
\label{tab:correlation}}
\tablecolumns{6}
\tablehead{
\colhead{} &
\multicolumn{2}{c}{$|l|$} &
\colhead{} &
\multicolumn{2}{c}{$|b|$} 
\\
\cline{2-3}
\cline{5-6}
\colhead{} &
\colhead{$\rho$} &
\colhead{$p$-value} & 
\colhead{} &
\colhead{$\rho$} & 
\colhead{$p$-value} 
}
\startdata
\kThalo{} & $0.02$ & $0.79$ & & $-0.24$ & $0.019$\\
\EMhalo{} & $-0.09$ & $0.39$ & & $-0.24$ & $0.012$\\
\Fehalo{} & $0.08$ & $0.53$ & & $0.03$ & $0.70$\\
\enddata
\end{deluxetable}

\begin{figure*}
\gridline{
	\leftfig{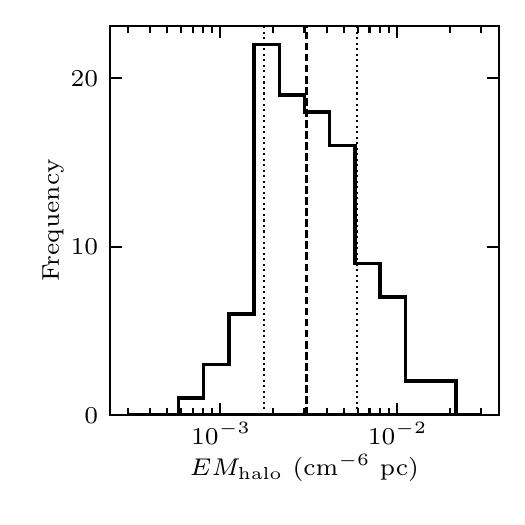}{0.32\textwidth}{}
	\fig{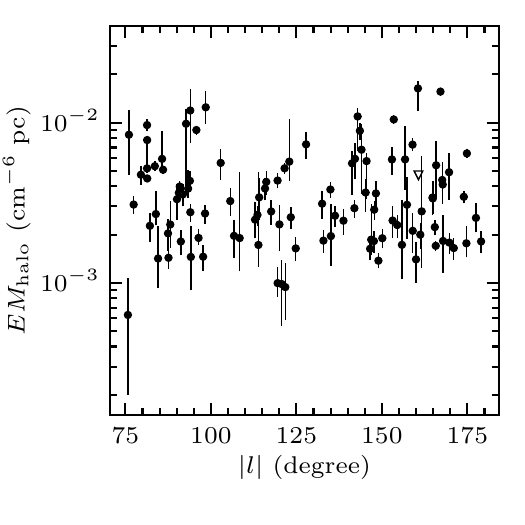}{0.32\textwidth}{}
	\rightfig{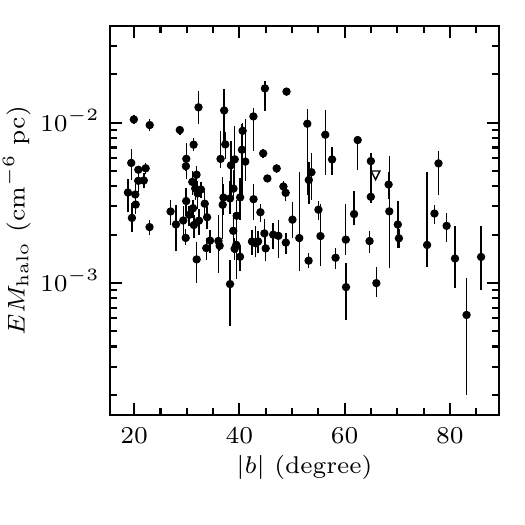}{0.32\textwidth}{}
}
\caption{Same as \Fig{fig:kThalo} but for \EMhalo{}.
The 3$\sigma$ upper limit is shown by the downward-pointing triangles in the middle and the right panels.
The upper limit is excluded in the left histogram. 
\label{fig:EMhalo}}
\end{figure*}

\Fig{fig:EMhalo} is the same as \Fig{fig:kThalo} but for \EMhalo{}.
The histogram of \EMhalo{} is spread over more than one order of magnitude;
the minimum is $0.6\times10^{-3}$\EMunit{}, the maximum is $16.4\times10^{-3}$\EMunit{}, and the median is $3.1\times10^{-3}$\EMunit{}.
As shown in \Tab{tab:correlation}, no significant correlation was found between \EMhalo{} and $|l|$, whereas a weak negative correlation of $\rho=-0.25$ was found between \EMhalo{} and $|b|$ with the $p$-value of 0.012.

\begin{figure*}
\gridline{
	\leftfig{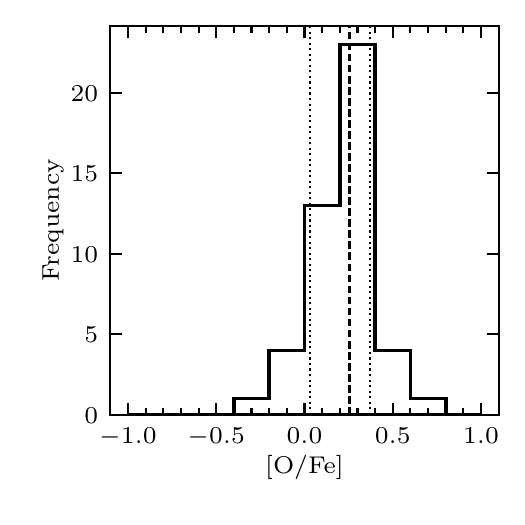}{0.32\textwidth}{}
	\fig{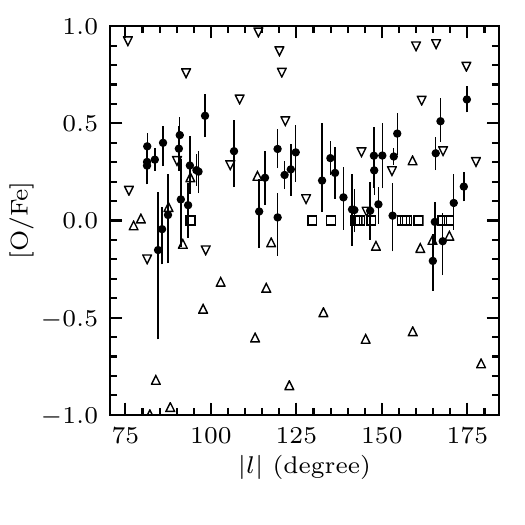}{0.32\textwidth}{}
	\rightfig{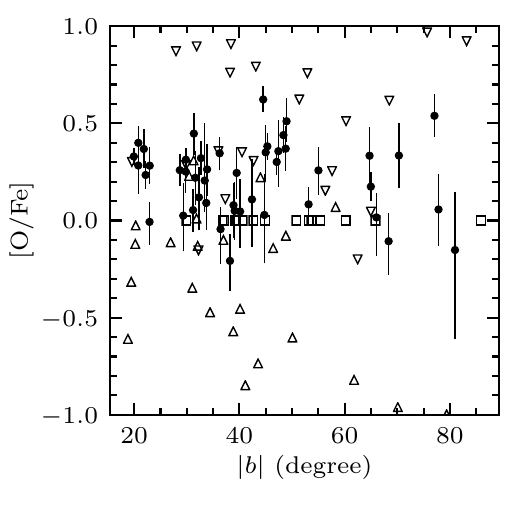}{0.32\textwidth}{}
}
\caption{Same as \Fig{fig:kThalo} but for \Fehalo{}.
In the middle and right panels, the 3$\sigma$ upper and lower limits are shown by the downward-pointing and upward-pointing triangles, respectively.
The fixed values are shown by the squares.
The upper and lower limits and fixed values are excluded from the left histogram. 
\label{fig:Fehalo}}
\end{figure*}

\Fig{fig:Fehalo} is the same as \Fig{fig:kThalo} but for \Fehalo{}.
Because \Fehalo{} is constrained in only 46 out of 107 regions, the histogram is drawn for those 46 fields.
The median is 0.25 and the 16--84th percentile range is 0.03--0.37.
We found no significant correlation between \Fehalo{} and the Galactic coordinates (\Tab{tab:correlation}).

The 68\% interval with the median of $N_\mathrm{CXB}$ is $9.0_{-1.4}^{+1.0}$ ~photons~cm$^{-2}$~s$^{-1}$~sr$^{-1}$~keV$^{-1}$.
The median value is $\sim10$\% lower than the value reported by \cite{2017ApJ...837...19C} but is within the systematic uncertainty of the different measurements \citep{2009A&amp;A...493..501M}.
The fluctuation of $N_\mathrm{CXB}$ is consistent with the cosmic variance expected in the Suzaku FoV ($\sim$15\%). 

The range of \EMlocal{} is 6.4--$33\times10^{-3}$\EMunit{} with a median of $18\times10^{-3}$\EMunit{}.
The surface brightness of the local emission component spans 1.3--$6.8\times10^{-12}$~erg~cm$^{-2}$~s$^{-1}$~deg$^{-2}$ in the 0.4--1.0~keV band.
These parameter ranges roughly agree with those obtained by previous observations with Suzaku and XMM-Newton \citep{2007PASJ...59S.141S,2007ApJ...658.1081G,2015ApJ...808...22H,2016ApJ...816...33U}

\subsection{Correlations between the parameters}
\label{sec:anal:correlation}
\begin{figure*}
\gridline{
	\leftfig{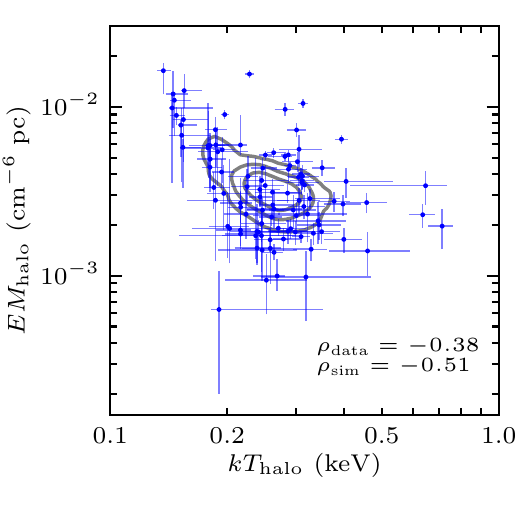}{0.32\textwidth}{}
	\fig{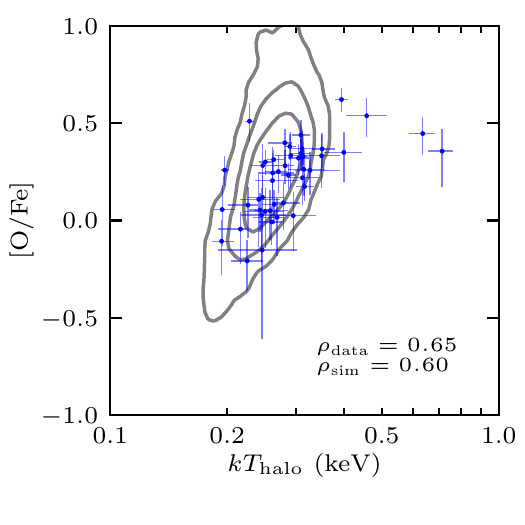}{0.32\textwidth}{}
	\rightfig{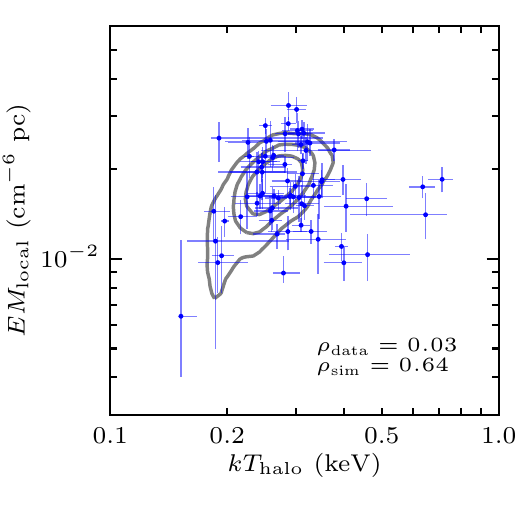}{0.32\textwidth}{}
}
\caption{
Scatter plots of \EMhalo{}, \Fehalo{}, and \EMlocal{} versus \kThalo{}.
The data of the upper and lower limits are not shown. 
The gray contours indicate the 68\%, 90\%, and 99\% ranges of the simulation results where the typical parameters and the exposure time were assumed.
\label{fig:correlation}}
\end{figure*}

Correlations between the parameters (\EMhalo{}, \Fehalo{}, and \EMlocal{} versus \kThalo{}) are shown in \Fig{fig:correlation}.
The corresponding Spearman correlation factors ($\rho_\mathrm{data}$) are also shown.
The \EMhalo{}--\kThalo{} and \Fehalo{}-\kThalo{} plots show negative and positive correlations, respectively, whereas the \EMlocal{}--\kThalo{} plot shows no correlation.

These correlations might be artifacts due to intrinsic correlations in the spectral model,
because, even if all the fields have the same true values, the obtained fitting parameters may have some correlations due to statistical uncertainties.
To investigate this effect, we created $10^{4}$ simulated spectra of XIS1 with a typical exposure time of 50~ks and  the median values of the parameters obtained in the previous section.
We derived the best-fit parameters from these mock spectra and created statistical contours on \Fig{fig:correlation}, which indicated the intrinsic correlations between the parameters.
The Spearman correlation factors for the simulated dataset ($\rho_\mathrm{sim}$) are also shown in \Fig{fig:correlation}.
 
In the plots of \EMhalo{}--\kThalo{} and \EMlocal{}-\kThalo{}, the scatter of the data points is larger than the contours derived from the simulation, suggesting that the observed scatters do not originate from intrinsic correlations. 
On the other hand, the scatter of the data in the \Fehalo{}--\kThalo{} plot agrees with the contours, suggesting that the observed correlation is likely artificial.

\section{Discussion} \label{sec:discus}
We obtained the temperatures, the emission measures, and the [O/Fe] abundances of the hot gas for the 107 lines-of-sight.
We compared our result with those of previous studies in \Sec{sec:discuss:comparison}.
The contamination from unresolved stellar sources was estimated in \Sec{sec:discuss:star}. 
We then examined the spatial distribution model with our emission measure data in \Sec{sec:discuss:distribution} and discussed the origin of the hot gaseous halo in \Sec{sec:discuss:origin}.
We also discussed the metallicity and the high-temperature regions in  \Sec{sec:discuss:metal} and \Sec{sec:discuss:hotcomp}, respectively.

\subsection{Comparison with previous studies }    
\label{sec:discuss:comparison}
\subsubsection{Previous Suzaku results}
\label{sec:discuss:comparisonSuzaku}

\citet{2009PASJ...61..805Y} (hereafter Y09) analyzed 13 Suzaku observations, 9 of which are also included in our dataset.
Before making a comparison with our result, we need to note the differences between their spectral model ("model2") and our model.
Y09 used the solar abundance of \cite{1989GeCoA..53..197A} and the old AtomDB version 1.3.1.
They fixed \EMlocal{} to $7\times10^{-3}$\EMunit{}, which is lower than our best-fit median value by a factor of 2.5.
The neon abundance is a free parameter in Y09 in contrast to the fixed solar value used in our analysis.
Their CXB is modeled by two broken power-law functions instead of a single power-law function.
They calculated the absorption column densities from \cite{1990ARA&A..28..215D}, which are slightly lower than those from \cite{2013MNRAS.431..394W}.

\begin{figure}
\epsscale{1.2}
\plotone{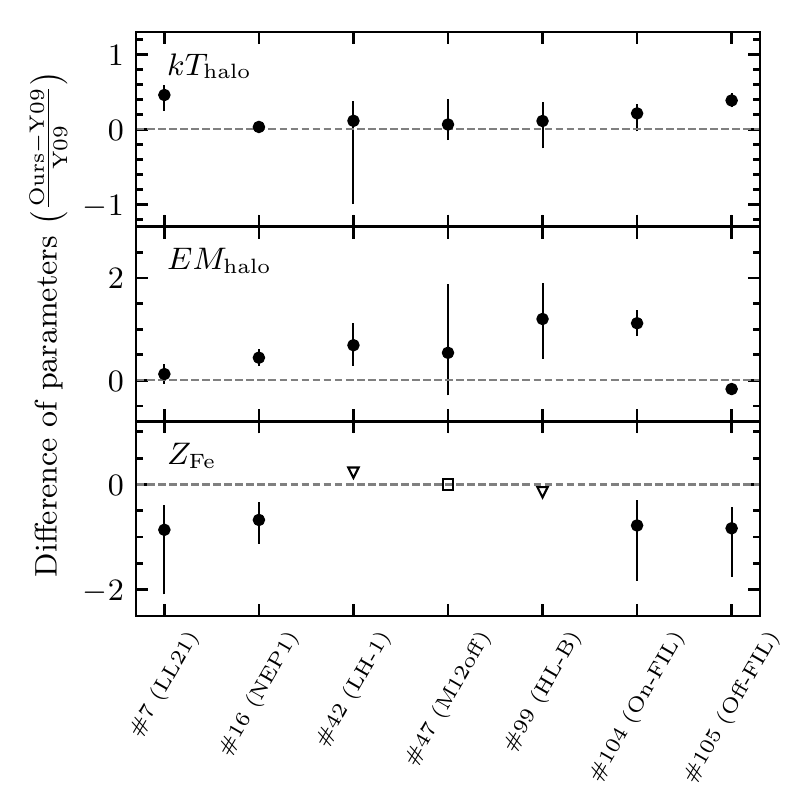}
\caption{Comparison of the best-fit parameters between our results and those of Y09 for the same observations. 
The region numbers in this paper are shown on the horizontal axis with the corresponding region names in Y09.
Statistical errors from the two results are added quadratically.
The downward-pointing triangles indicate the upper limits. 
The square indicates that the parameter is fixed in both this study and Y09.
\label{fig:compareY09}}
\end{figure}

\Fig{fig:compareY09} compares our results to those of Y09  using the nine overlapped regions.
Y09 analyzed the two Lockman Hole observations (LH-1 and LH-2) and the two north ecliptic pole observations (NEP1 and NEP2) separately;
however, we showed only a comparison with the LH-1 and NEP1 results in \Fig{fig:compareY09} because the same line-of-sight data were simultaneously fitted in our analysis. 
We found that for our results \kThalo{} is $\sim20$\% higher, \EMhalo{} is $\sim60$\% higher, and $Z_\mathrm{Fe}$ is $80$\% lower compared to the Y09 results on average.

These discrepancies results from the model differences described below.
First is the lower \EMlocal{} in Y09 compared to our best-fit values.
The lower \EMlocal{} deceases \kThalo{} to compensate for the \ion{O}{7} line flux.
This tendency is  illustrated in the right panel of \Fig{fig:correlation}.
Second is the difference in the solar abundance.
Y09 used the solar abundance of \cite{1989GeCoA..53..197A}, in which the oxygen abundance was 41\% higher than that shown in \cite{2009LanB...4B...44L}.
The oxygen abundance directly affects the flux (and therefore \EMhalo{}) of the hot gaseous halo model because emissions from oxygen, including the radiative recombination continua, dominate the 0.4--1.0~keV flux of a hot gas with $kT\sim0.26$~keV.
Third is the difference in the AtomDB versions.
The emissivities of the Fe L-shell lines in the 0.7--0.9~keV band for AtomDB 1.3.1 were $\sim60$\% lower than those for AtomDB 3.0.9.
That leads to higher iron abundances in Y09 compared to our results.
Forth is the treatment of the neon abundance.
The free neon abundance has a slight effect on \kThalo{} and \EMhalo{}.
When we re-analyzed our data with the same settings as Y09 for the above four points, we obtained the results consistent with those of Y09.
We also confirmed that the differences in the CXB model and the absorption column density hardly affect the results. 

The parameter differences between this study and Y09's study are not considered to be systematic uncertainties for the following reasons.
First, there is no incentive to fix \EMlocal{} to a certain value, considering the one order of magnitude flux variation in the local component found by other observations \citep[e.g.,][]{2015ApJ...808...22H,2017ApJ...834...33L}.
Second, using an up-to-date database of the solar abundance and the atomic database provides the current best estimates of the parameters.
In particular, the emissivities of the strong Fe L-shell lines have been calibrated with grating spectrometer observations over the past decade.  
Third, fixing the neon abundance to the solar value as same as the oxygen is physically motivated as both neon and oxygen are primarily synthesized by core-collapse supernovae, and therefore they are likely to have the same abundance relative to the solar values.
Indeed, the abundances of oxygen and neon relative to the solar values are consistent with each other in the intracluster medium \citep[e.g.,][]{2016A&amp;A...595A.126M}.

\subsubsection{Previous XMM-Newton results}
\label{sec:discuss:comparisonXMM}
\citet{2013ApJ...773...92H} (hereafter HS13) analyzed 110 lines-of-sight out of the Galactic plane ($|b|>30\arcdeg$) using the XMM-Newton observations.
They derived temperatures and emission measures from the 0.4--5.0~keV spectral modeling assuming  solar metallicity for the hot gas.

\begin{figure}
\epsscale{1.2}
\plotone{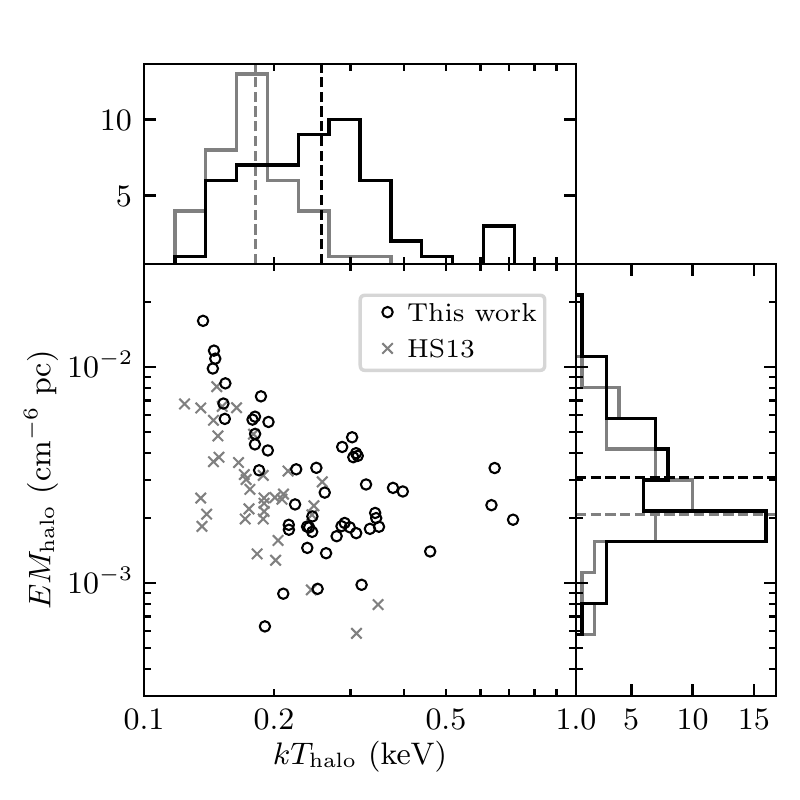}
\caption{Scatter plot of \EMhalo{} versus \kThalo{} for this study (black) and those of H13 (gray).
We only show the data at $75\arcdeg<l<285\arcdeg$ and $|b|>30\arcdeg$ to match the analyzed sky regions.
Error bars are not shown to simplify the plot.
The histograms on the top and right sides of the scatter plot are the distributions of \kThalo{} and \EMhalo{}, respectively, with the median shown by the dashed lines.
\label{fig:compareHS13}}
\end{figure}

We compared \kThalo{} and \EMhalo{} from our study with those of HS13 (\Fig{fig:compareHS13}).
In this plot, we only show the data at $75\arcdeg<l<285\arcdeg$ and $|b|>30\arcdeg$, areas that both this study and HS13 analyzed.
The scatter plot shows a similar trend between the two.
However,  the median temperature from our result is $\sim0.1$~keV higher than that of HS13, and the median emission measure from our result is $\sim50$\% higher than that of HS13. 

The shift in the median \kThalo{} is likely due to the differences in \EMlocal{}.
We allowed \EMlocal{} to vary,
while HS13 fixed \EMlocal{} according to the count rates of the ROSAT R12 band obtained from the  shadowing observations of nearby dark clouds \citep{2000ApJS..128..171S} because it is difficult to determine \EMlocal{} from the XMM-Newton spectrum itself due to the heavy contamination of the soft proton background below 1~keV.
The median ROSAT count rate in the HS13 analysis is $\sim600$~counts~s$^{-1}$~arcmin$^{-1}$.
Assuming a temperature of 0.1~keV and solar metallicity, this count rate can be converted to  an\EMlocal{} of $\sim4\times10^{-3}$\EMunit{}, which is lower than our median value of $15\times10^{-3}$\EMunit{} by a factor of $\sim4$.
Lower \EMlocal{} leads to lower \kThalo{} as shown by the contours in the right panel of \Fig{fig:correlation}; if we fixed \EMlocal{} to $4\times10^{-3}$\EMunit{}, the median \kThalo{} decreases to 0.18~keV and becomes consistent with that of HS13 but the fitting statistics become considerably worse.
The result indicates that the extrapolation of the ROSAT R12 band (0.11--0.28~keV) flux to the  analysis energy band (0.4--5.0~keV) has systematic uncertainties due to the different  contributions of the SWCX emission between those two bands and/or the different solar activity between the ROSAT era and the Suzaku/XMM-Newton era. 

The shift in the median \EMhalo{} is caused by the difference in the solar abundance.
As shown in the case of Y09, HS13 used the solar abundance of \cite{1989GeCoA..53..197A}.
The 41\% higher oxygen abundance in \cite{1989GeCoA..53..197A} compared to that in\cite{2009LanB...4B...44L} increases the flux of the hot gaseous halo model by $\sim40$\%.
This explains the difference in \EMhalo{} between our results and those of H13. 

\subsection{Contamination of unresolved stellar sources}
\label{sec:discuss:star}
\cite{2001ApJ...554..684K} estimated the contribution of unresolved stellar sources to the soft X-ray background flux measured with ROSAT, and concluded that that is negligible at least for $|b|>30\arcdeg$.
\cite{2009PASJ...61..805Y} calculated the flux of unresolved dM stars assuming the stellar distribution model and found that the integrated flux is lower than the observed flux by a factor of 5.
Even though minor contributions of stellar sources to the soft X-ray background were shown in  previous studies, we re-evaluated the possible contamination of stellar sources at low Galactic latitudes using recent observations of stellar sources.  

According to the $\log N$-$\log S$ plot of active coronae reported by \cite{2013A&A...553A..12N}, the integrated surface brightness of the unresolved stars below a flux of $5\times10^{-14}$~erg~cm$^{-2}$~s$^{-1}$ is $\sim0.5\times10^{-12}$~erg~cm$^{-2}$~s$^{-1}$~deg$^{-2}$.
That is $<20$\% of the surface brightness of the hot gaseous halo at $15\arcdeg<|b|<20\arcdeg$.
Therefore, the contribution of unresolved stellar sources to the observed flux is not significant.

\subsection{Spatial distribution of the hot gas}
\label{sec:discuss:distribution}

\begin{figure*}
\gridline{
	\fig{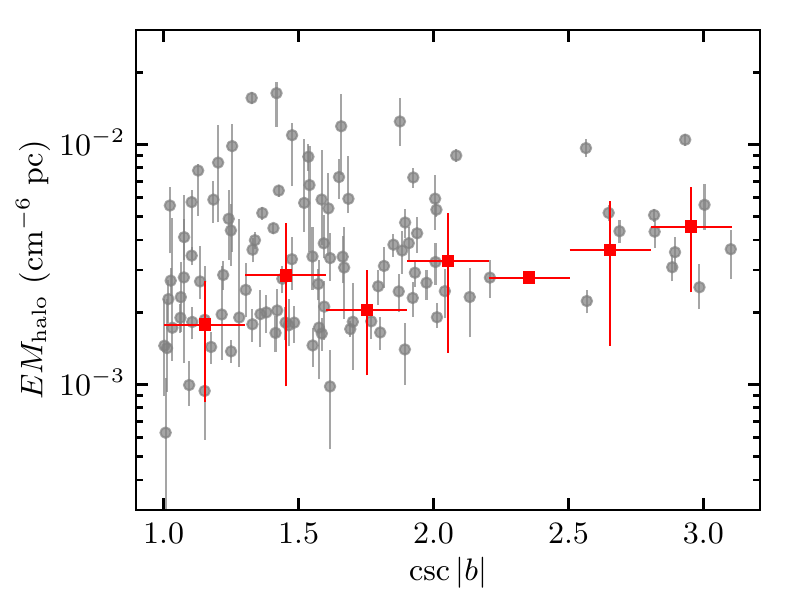}{0.45\textwidth}{}
	\fig{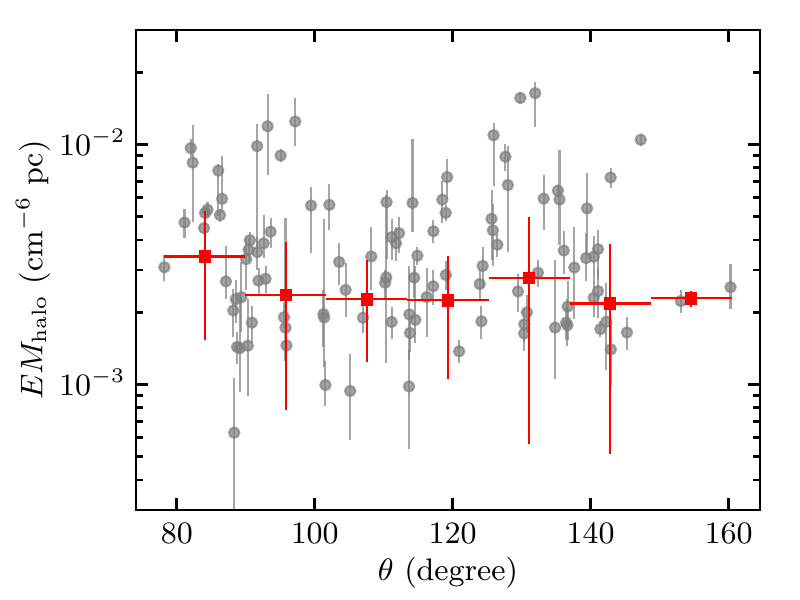}{0.45\textwidth}{}
}
\caption{Emission measures of the hot gaseous halo component along $\csc |b|$ (left) and $\theta = \arccos(\cos l\cos b)$  (right).
The gray data points are the emission measures for individual Suzaku lines-of-sight, whereas the red data points are the inverse squared-error weighted means calculated from the intervals shown in the horizontal red bars. 
The vertical red bars indicate the corresponding inverse squared-error weighted standard deviations.
\label{fig:EMcscbangFC}}
\end{figure*}

Two types of density distribution models for the hot gaseous halo have been proposed.
One is a disk-like morphology suggested by the combined analysis of emission and absorption line measurements \citep[e.g.,][]{2009ApJ...690..143Y}.
The other is a spherical distribution model, in particular, the modified $\beta$-model constructed by \cite{2013ApJ...770..118M,2015ApJ...800...14M}.
We compared these two models with our emission measure data.

Spatial correlations between \EMhalo{} and the Galactic coordinates are key to distinguishing the models.
The disk-like morphology predicts that \EMhalo{} is proportional to $\csc|b|$ ($=1/\sin|b|$). 
Conversely, the spherical distribution model predicts decreasing \EMhalo{} with increasing  angle from the Galactic Center ($\theta = \arccos(\cos l\cos b)$).
We examined these points in \Fig{fig:EMcscbangFC}.
The binned data (red crosses) are also shown in the figure to smooth out the large scatter of the data points.
A positive correlation is shown in the left panel (\EMhalo{}-$\csc|b|$ plot), whereas no clear correlation is shown in the right panel (\EMhalo{}-$\theta$ plot).
Therefore, a disk-like morphology is qualitatively favored.

To perform quantitative analyses, we formulated the models as follows. 
According to \cite{2017ApJ...849..105L}, the disk model ($n_\mathrm{disk}$) is parameterized by the scale length ($R_0$) and the scale height ($z_0$) such that 
\begin{equation}
n_\mathrm{disk}(R,z)  = n_\mathrm{0}\exp\left({-\frac{R}{R_0}}\right)\exp\left({-\frac{z}{z_0}}\right),
\end{equation}
where $R$ is the distance from the Galactic Center projected onto the Galactic plane, $z$ is the vertical height from the Galactic plane, and $n_\mathrm{0}$ is the normalization factor corresponding to the number density at the Galactic Center.
The  spherical distribution model (modified $\beta$ model) used by \cite{2015ApJ...800...14M} is described as 
\begin{equation}
n_\mathrm{sphe}(r) = n_{c}\left(\frac{r}{r_{c}}\right)^{-3\beta},
\end{equation}
where $n_\mathrm{sphe}$ is the number density, $r$ is the distance from the Galactic Center, $n_{c}$ is the core density,  $r_c$ is the core radius, and $\beta$ is the slope of the profile.
Assuming a line-of-sight distance from the Sun ($s$), $R$, $z$, and $r$ are described as a function of the Galactic coordinates:
\begin{eqnarray}
R(l,b,s) &=& \sqrt{D_{\sun}^2 +(s \cos b)^2 - 2 D_{\sun} s \cos b \cos l},  \\
z(b,s) &=& s\sin b, \\
r(l,b,s) &=& \sqrt{R(l,b,s)^2 + z(l,b,s)^2},
\end{eqnarray}
where $D_{\sun}$ is the distance between the Sun and the Galactic Center (8~kpc).
We then derive the emission measures predicted by these density models at a certain line-of-sight as 
\begin{eqnarray}
EM_\mathrm{disk}(l,b) &=& \int_{0}^{s_\mathrm{max}}n^2_\mathrm{disk}(R(l,b,s),z(b,s)) ds, \\
EM_\mathrm{sphe}(l,b) &=& \int_{0}^{s_\mathrm{max}}n^2_\mathrm{sphe}\left(r(l,b,s)\right) ds,
\end{eqnarray}
where $s_\mathrm{max}$ is the maximum path length of the integration.
We assumed a $s_\mathrm{max}$ of 100~kpc in the following discussion.
Values of $s_\mathrm{max}$ larger than 100~kpc did not affect the results.

\begin{figure}
\epsscale{1.2}
\plotone{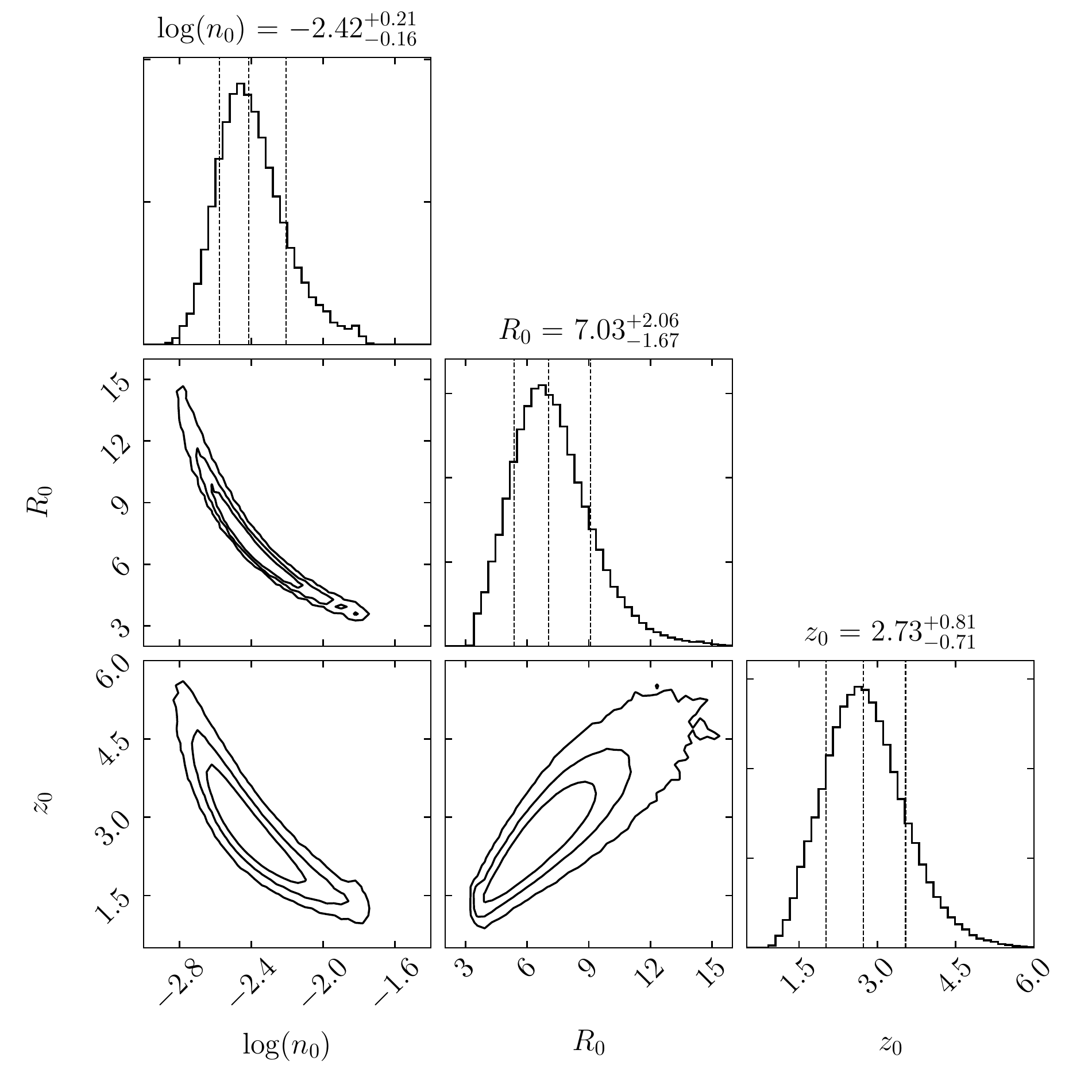}
\caption{Posterior probability distributions of the disk-model parameters derived from the fitting with the MCMC simulations. The vertical dotted lines of each histogram indicate the 16th, 50th, and 84th percentiles. Contours indicate the 68\%, 90\%, and 99\% levels.
\label{fig:triangledisk}}
\end{figure}

\begin{figure*}
\gridline{
	\fig{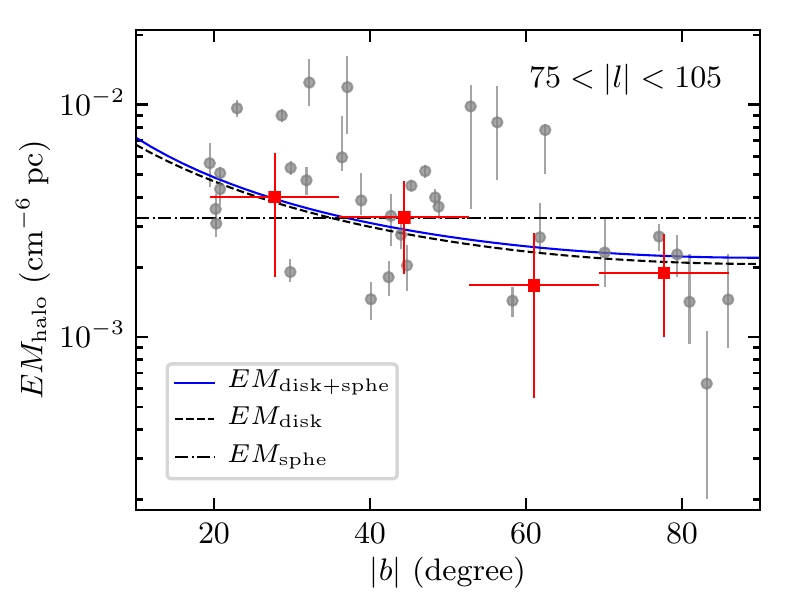}{0.45\textwidth}{(a)}
	\fig{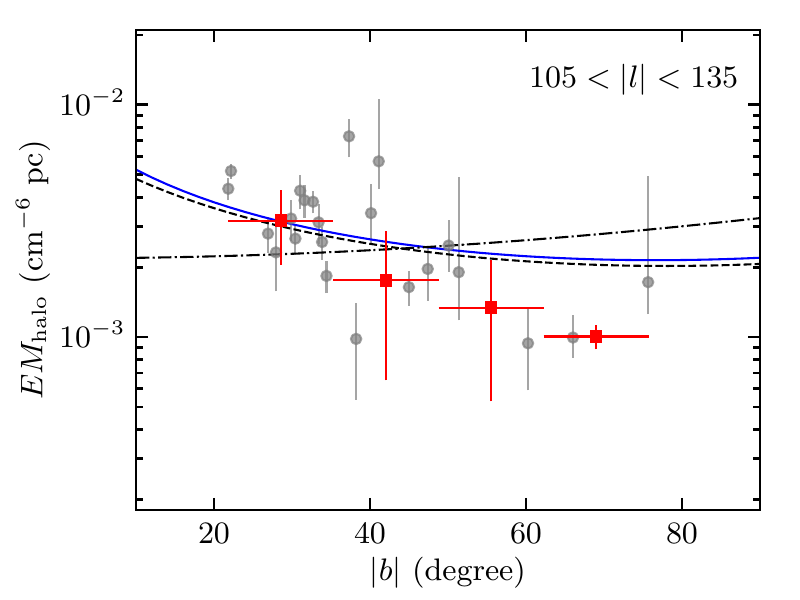}{0.45\textwidth}{(b)}
}
\gridline{
	\fig{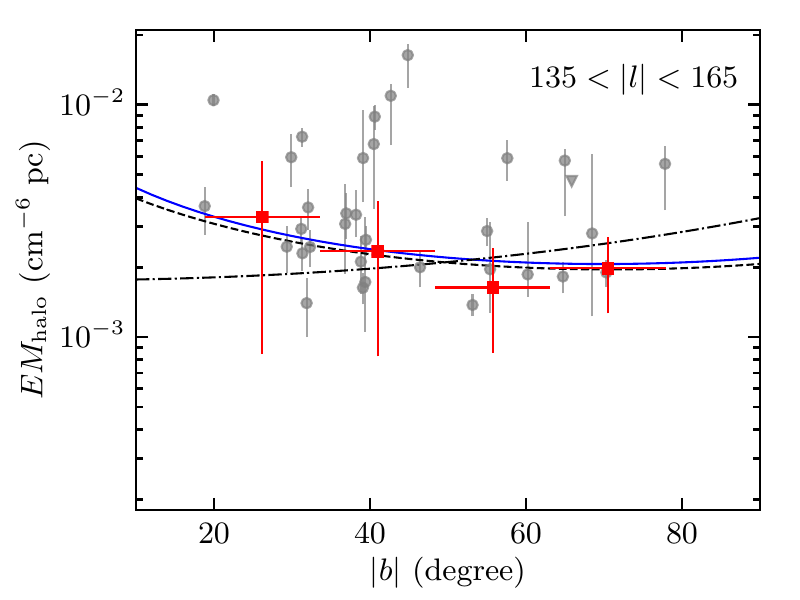}{0.45\textwidth}{(c)}
	\fig{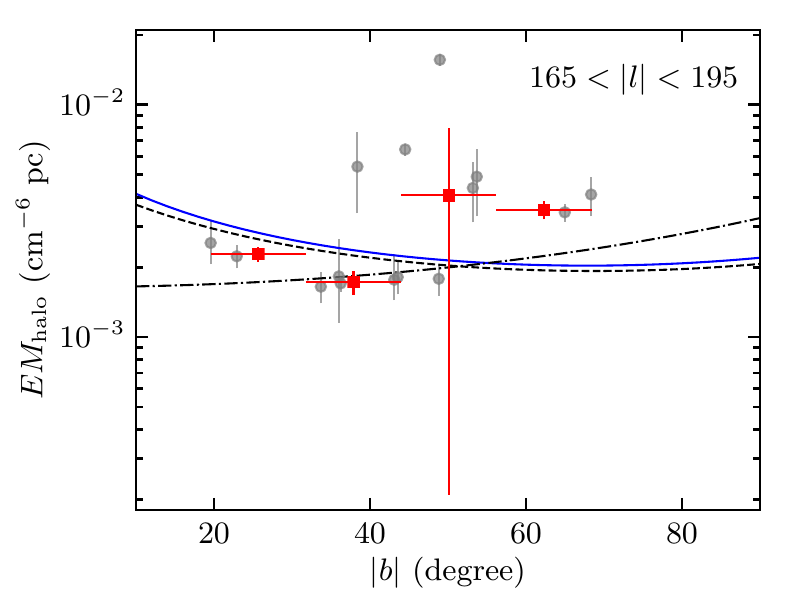}{0.45\textwidth}{(d)}
          }
\caption{Emission measures of the hot gaseous halo versus the absolute value of the Galactic latitude.
Each panel shows a different range of the Galactic longitude: (a) $|l| = 90\arcdeg\pm15\arcdeg$, (b) $|l| = 120\arcdeg\pm15\arcdeg$, (c) $|l| = 150\arcdeg\pm15\arcdeg$, and (d) $|l| = 180\arcdeg\pm15\arcdeg$.
The fitted disk and spherical models are shown by the dashed and dot-dashed curves, respectively, and the full (uppermost) lines are their composite.
The red data points are the inverse squared-error weighted means calculated from the intervals shown by the horizontal bars with weighted standard deviations shown by the vertical bars. 
\label{fig:fitEM}}
\end{figure*}

First, we fitted the $EM_\mathrm{disk}$ model to the data using the Markov chain Monte Carlo (MCMC) package \texttt{emcee} \citep{2013PASP..125..306F}.
The maximum likelihood estimator was constructed from the $\chi^2$ values. 
We ran $10^{7}$ steps with an ensemble of 100 walkers.
We confirmed that the autocorrelation times of each parameter were shorter than the step numbers by a factor of $>10$.
Posterior distributions were constructed from the last $10^{5}$ steps (hence $10^{7}$ samples).
\Fig{fig:triangledisk} shows the resulting posterior distribution of the $EM_\mathrm{disk}$ model.
The medians of the parameters are shown \Tab{tab:fitEM}.
The quoted uncertainties are  the 16th to 84th percentiles.
The dashed curves in \Fig{fig:fitEM} show representatives of the $EM_\mathrm{disk}$ model at $|l|=90\arcdeg$, $120\arcdeg$, $150\arcdeg$, and $180\arcdeg$.
Observed emission measures in the corresponding $|l|$ ranges are also shown by the gray points.
The model approximates the observed data even though  a large scatter ($\approx40$\%) of the data around the model is present.
To smooth out the possible intrinsic scatter of the data, we also show the binned data in \Fig{fig:fitEM} with the red crosses.
The binned data roughly agree with the $EM_\mathrm{disk}$ model.
The obtained $n_0$ and $z_0$  are consistent with those derived from previous studies toward LMC~X-3, PKS~2155--204, and Mrk~421, where $n_0 = 1$--$5\times10^{-3}$~cm$^{-3}$ and $z_0 = 2$--9~kpc \citep{2009ApJ...690..143Y,2010PASJ...62..723H,2014PASJ...66...83S}.

\begin{deluxetable*}{cccccccc}
\tablecaption{Fitting results of the density distribution models. \label{tab:fitEM}}
\tablehead{
& \multicolumn{3}{c}{Parameters of the disk model} && \multicolumn{3}{c}{Parameters of the spherical model }  \\
\cline{2-4}\cline{6-8}
\colhead{Model} &  \colhead{$n_0$~($10^{-3}$~cm$^{-3}$)} & \colhead{$R_0$~(kpc)} & \colhead{$z_0$~(kpc)}&& \colhead{$n_c$~($10^{-3}$~cm$^{-3}$)}& \colhead{$r_c$~(kpc)} & \colhead{$\beta$} 
}
\startdata
$EM_\mathrm{disk}$ & $3.8^{+2.2}_{-1.2}$ & $7.0^{+2.1}_{-1.7}$ & $2.7^{+0.8}_{-0.7}$ && \nodata & \nodata & \nodata \\
$EM_\mathrm{sphe}$ & \nodata & \nodata & \nodata && $4.3^{+0.1}_{-0.1}$ & 2.4 (fixed) & 0.51 (fixed) \\
$EM_\mathrm{disk+sphe}$ & $3.7^{+0.4}_{-0.4}$ & 7.0 (fixed) & $1.8^{+0.7}_{-0.8}$ && $1.2^{+0.8}_{-0.8}$ & 2.4 (fixed) & 0.51 (fixed) \\ 
\enddata
\tablecomments{Uncertainties are the 16th to 84th percentile ranges of the posterior distributions}
\end{deluxetable*}

Then, we fitted the $EM_\mathrm{sphe}$ model in the same manner as the above $EM_\mathrm{disk}$ fitting.
Because $r_c$ and $\beta$ were not well constrained in our fitting, we fixed them to 2.4~kpc and 0.51, respectively, according to the results of \cite{2017ApJ...849..105L}.
The fitted parameters are shown in \Tab{tab:fitEM}, and the representative mode curves (dot-dashed curves) are shown in \Fig{fig:fitEM}.
In contrast to the $EM_\mathrm{disk}$ model, the $EM_\mathrm{sphe}$ model increases with increasing $|b|$ and therefore is not in line with the tendency of the data especially at $75\arcdeg < |l| < 105\arcdeg$ and $105\arcdeg < |l| < 135\arcdeg$.
The obtained $n_c$ is consistent with the value shown in \cite{2015ApJ...800...14M} and is approximately a half of that in \cite{2017ApJ...849..105L}.

\begin{figure}
\epsscale{1.2}
\plotone{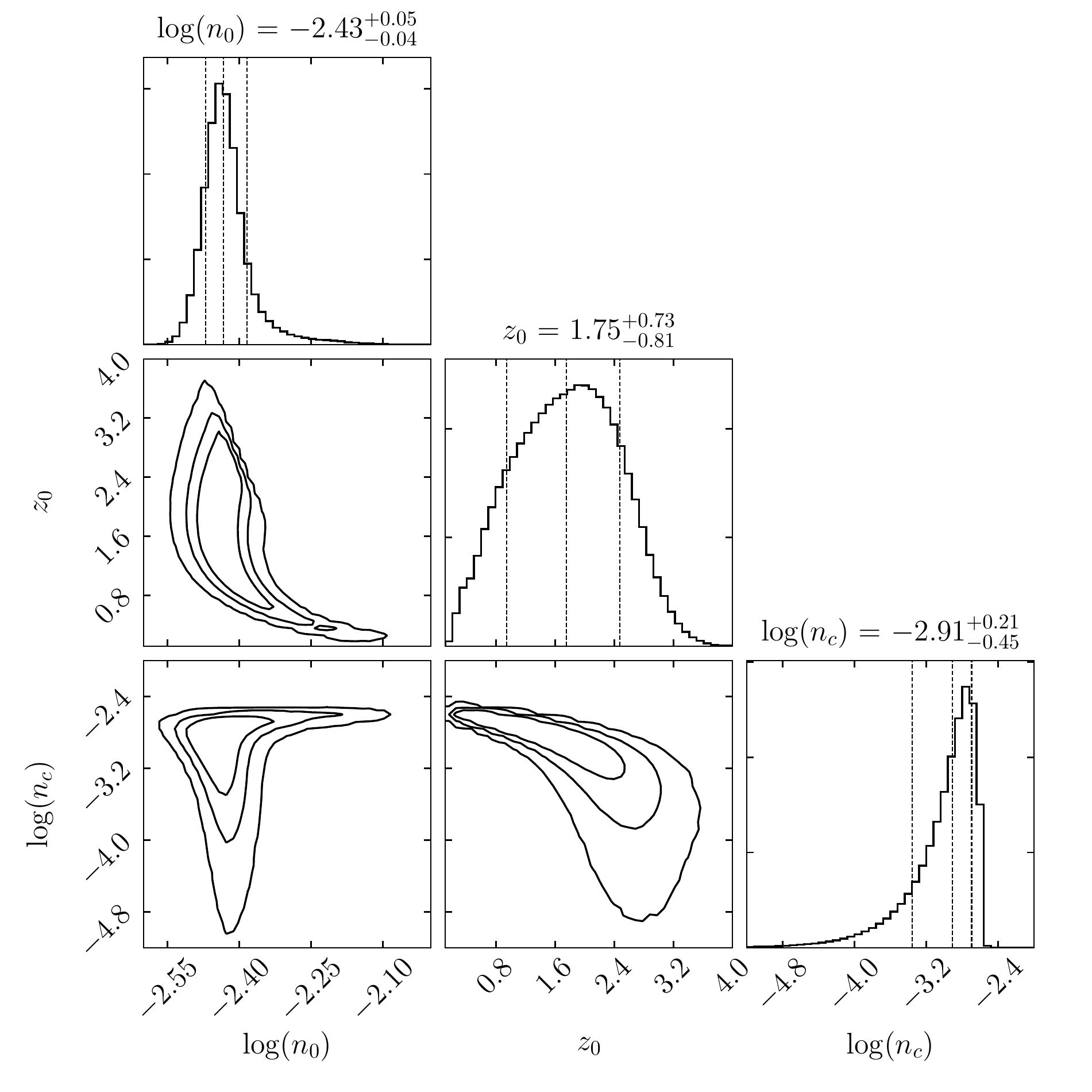}
\caption{Same as \Fig{fig:triangledisk} but for the $EM_\mathrm{disk+sphe}$ model. 
\label{fig:trianglecomp}}
\end{figure}

Finally, we constructed a composite of the disk and the spherical models where the density and the emission measure are described as 
\begin{equation}
n_\mathrm{disk+sphe}  = n_\mathrm{disk} + n_\mathrm{sphe}
\end{equation}
and 
\begin{equation}
EM_\mathrm{disk+sphe} = \int_0^{s_\mathrm{max}} n_\mathrm{disk+sphe}^2 ds,
\end{equation}
respectively.
In this composite model, we fixed $r_c$ and $\beta$ as in the fitting of the $EM_\mathrm{sphe}$ model.
In addition, $R_0$ was fixed to 7.0~kpc, which was obtained from the $EM_\mathrm{disk}$ model fitting, because this parameter was not well constrained in the composite model fitting.
The fitting with the MCMC simulation gives the posterior distributions shown in \Fig{fig:trianglecomp} and the parameter ranges summarized in \Tab{tab:fitEM}.
The fitted parameters of the disk-model component are consistent with those obtained from the $EM_\mathrm{disk}$ model fitting, while the normalization of the spherical-model component is lower than that obtained from the $EM_\mathrm{sphe}$ model fitting by a factor of $\sim4$. 
The blue curves in \Fig{fig:fitEM} are representatives of the $EM_\mathrm{disk+sphe}$ model at $|l|=90\arcdeg$, $120\arcdeg$, $150\arcdeg$, and $180\arcdeg$.
As shown in this figure, the $EM_\mathrm{disk+sphe}$ model is nearly the same as the $EM_\mathrm{disk}$ model and the contribution of the spherical-model component is minor.

A similar composite model was also examined by \cite{2017ApJ...849..105L} using the emission line data of XMM-Newton.
For comparison, we calculated the model densities at the solar neighborhood;
$n_\mathrm{disk, \sun}$ was calculated from the disk model at $R=8$~kpc and $z=0$~kpc, and  $n_\mathrm{sphe, \sun}$ was calculated from the spherical model at $r=8$~kpc.
These values are shown in \Tab{tab:density}.
Both results indicate that the density of the disk model is higher than that of the spherical model  in the solar neighborhood.
The quantitative difference likely reflects systematic uncertainties between the different analysis methods.
For example, \cite{2017ApJ...849..105L} assumed a constant temperature of $2 \times 10^6$~K; however, our spectroscopic results show that the median temperature is $3 \times 10^6$~K with $\sim30$\% fluctuations.

\begin{deluxetable}{lccc}
\tabletypesize{\scriptsize}
\tablecaption{Model densities in the solar neighborhood. \label{tab:density}}
\tablehead{
\colhead{Model} & 
\colhead{$n_\mathrm{disk,\sun}$\tablenotemark{a}} & 
\colhead{$n_\mathrm{sphe,\sun}$\tablenotemark{b}} & 
\colhead{ $\frac{n_\mathrm{disk,\sun}}{n_\mathrm{sphe,\sun}}$ } \\
\colhead{}&
\colhead{($10^{-3}$~cm$^{-3}$)}&
\colhead{($10^{-3}$~cm$^{-3}$)}&
\colhead{}
} 
\startdata
This work & 1.2 & 0.2 & 6.1\\
\cite{2017ApJ...849..105L} & 2.5 & 1.2 & 2.1 \\
\enddata
\tablenotetext{a}{Density of the disk component at $R=8$~kpc and $z=0$~kpc.} 
\tablenotetext{b}{Density of the spherical component at $r=8$~kpc. }
\end{deluxetable}

\begin{figure}
\epsscale{1.2}
\plotone{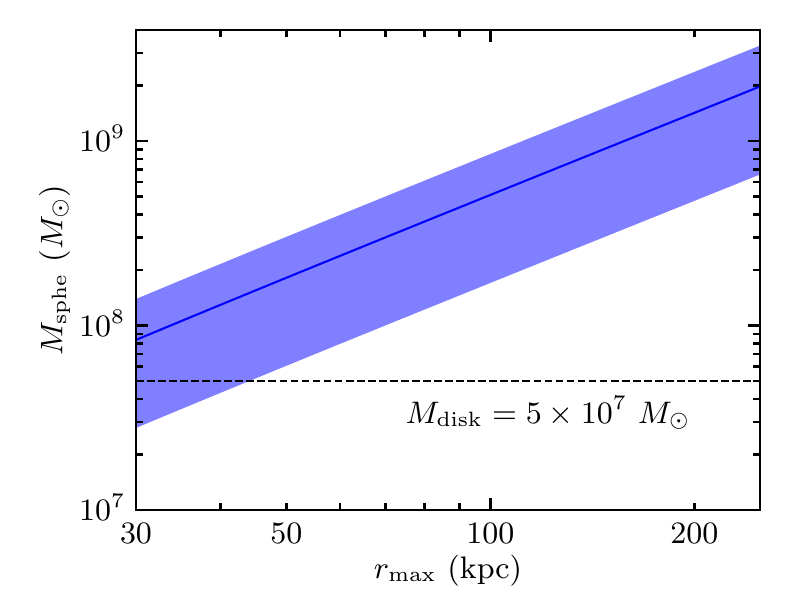}
\caption{Integrated mass of the spherical component as a function of the assumed $r_\mathrm{max}$.
The blue-hatched area shows the uncertainty obtained from the fitting.
The estimated $M_\mathrm{disk}$ is shown as the horizontal dashed line. 
Solar metallicity is assumed for both $M_\mathrm{sphe}$ and $M_\mathrm{disk}$.
\label{fig:mass}}
\end{figure}

The fitting with our composite model suggests that the observed X-ray emissions primarily originate in the disk component rather than in the spherical component.
However, the contribution to the mass of the gaseous halo has the opposite trend.
The total mass of the disk model component is 
\begin{eqnarray}
M_\mathrm{disk} &=& \int_{0}^{z_\mathrm{max}}\int_{0}^{R_\mathrm{max}}\frac{\mu m_p n_\mathrm{disk}(R,z)}{Z} 2\pi RdRdz \nonumber \\ 
&=& 5 \times 10^7 \left(\frac{Z}{Z_{\sun}}\right)^{-1} \mathrm{M}_{\sun}, \label{eq:massdisk}
\end{eqnarray}
where $\mu$ is the mean atomic weight of 0.61, $m_p$ is the proton mass, $Z$ is the metallicity of the gas, and both $R_\mathrm{max}$ and $z_\mathrm{max}$ are assumed to be 30~kpc.
Larger $R_\mathrm{max}$ and $z_\mathrm{max}$ do not affect the resulting mass.
On the other hand, the total mass of the spherical model component is described as
\begin{eqnarray}
M_\mathrm{sphe} &=& \int_{0}^{r_\mathrm{max}}\frac{\mu m_p n_\mathrm{sphe}(r)}{Z} 4 \pi r^2 dr \nonumber \\
&=& 2 \times 10^{9} \left(\frac{Z}{Z_{\sun}}\right)^{-1} \mathrm{M}_{\sun}, \label{eq:masssphe}
\end{eqnarray}
where $r_\mathrm{max}$ is assumed to be 250~kpc, which is the viral radius of our Galaxy.
As shown in \Fig{fig:mass}, even when $r_\mathrm{max}$ is $\sim30$~kpc, $M_\mathrm{sphe}$ is comparable to $M_\mathrm{disk}$.
Note that the extended spherical hot gas cannot explain the missing baryons in the MW ($\sim10^{11}$~M$_{\sun}$) even taking into account a low metallicity of $Z \sim 0.3Z_{\sun}$.

The smaller contribution of the spherical component to the X-ray emissions, despite its significant mass contribution, is caused by its low density because the X-ray flux of the diffuse hot gas is $\propto n^2$ and is biased toward high-density regions. 
To constrain the parameters of the spherical component, a large number of samples of absorption line measurements are necessary. 

 
\subsection{Origin of the hot gaseous halo}
\label{sec:discuss:origin}
Our X-ray emission data reveal the existence of a disk-like hot gas.
However, a more extended hot gas region is proposed by other indirect observations such as the pressure confinement of high velocity clouds in the MW halo \citep[e.g.,][]{2005ApJ...630..332F} and the ram-pressure stripping of local dwarf galaxies \citep[e.g.,][]{2009ApJ...696..385G}.
Therefore, we consider that the hot gaseous halo consists of a disk-like component and an extended spherical component.

A hot gas with a disk-like morphology is expected from stellar feedback in the MW disk; this is the so-called Galactic fountain model \citep[e.g.,][]{Shapiro_1976,1989ApJ...345..372N}.
The scale height we obtained ($\sim2$~kpc) is much smaller than that calculated from the assumption of hydrostatic equilibrium between the Galactic gravitational potential and the pressure gradient of a hot gas with a constant temperature of 0.26~keV (10--20~kpc). 
This indicates that the disk-like hot gas is not in hydrostatic equilibrium.
Indeed, numerical simulations of stellar feedback show a steep gradient of the hot gas density at $z \lesssim 1$~kpc, which is the launching site of hot gases generated by multiple supernovae in the galactic disk \citep{Hill_2012,2018ApJ...853..173K}.
The large scatter of the \EMhalo{} around the model is also naturally explained by the stellar feedback model.

One problem with the stellar feedback model is that it underpredicts the hot gas density (and therefore the X-ray flux) as reported by \cite{2015ApJ...800..102H}.
As pointed out by the authors, considering a spherically distributed hot gas and/or other driving mechanisms such as cosmic-ray driven outflows would mitigate the discrepancy between the observations and the numerical simulations.     

\subsection{The metal abundance of the hot gaseous halo}
\label{sec:discuss:metal}
For the first time, we derived a median \Fehalo{} of 0.25 using 46 lines-of-sight.
Even though this is subject to future updates of the atomic database and/or high resolution spectroscopy that resolves the Fe L-shell lines, this value is currently the best estimate with the latest databases.

The abundance ratio of \Fehalo{} provides complementary information concerning the origin of the hot gaseous halo.
Recent systematic observations of clusters of galaxies show that the ratio of $\alpha$-elements to Fe is consistent with the solar value in the intracluster medium \citep{2007A&A...462..953M,2016A&amp;A...595A.126M}.
This trend holds even at cluster outskirts, where the metallicity is as low as $\sim0.2$~$Z_\sun$ \citep{2015ApJ...811L..25S}.
Therefore, the intergalactic medium also likely holds [O/Fe] abundance ratio of the solar value. 
On the other hand, chemical composition of the outflowing hot gas from the MW disk reflects the recent rate of  core-collapse supernova (SN$_\mathrm{cc}$) to  type Ia supernova (SN$_\mathrm{Ia}$) in the MW because the cooling time of the hot gas is $\approx1$~Gyr.
Because the estimated SN$_\mathrm{cc}$ to SN$_\mathrm{Ia}$ rate for the recent MW is $\approx 5$ \citep{Li_2011},
 [O/Fe] is expected to be $\sim$0.17 according to metal yields of SN$_\mathrm{cc}$ and SN$_\mathrm{Ia}$ described in \cite{2006ApJ...653.1145K}.
The observed [O/Fe] roughly agrees with the above simple estimation, even though the actual abundance ratio is also affected by the mass loading factor which is highly uncertain. 
Therefore, it supports the stellar feedback scenario for the X-ray emitting hot gas rather than  accretion from the intergalactic medium.

\subsection{High-temperature regions}
\label{sec:discuss:hotcomp}
We found six lines-of-sight (22, 53, 70, 72, 75, and 94) that have temperatures of $>0.4$~keV, which is higher than the typical temperature range of 0.19--0.32~keV (\Fig{fig:highTspectra}).
These high-temperature regions are not concentrated in a specific sky region but are distributed randomly (\Fig{fig:kThalo}).
Such a high-temperature region was also reported by \cite{2013ApJ...773...92H} at ($l$, $b$) = ($237\fdg924$, $-54\fdg594$).

The origin of these high-temperature regions is still unclear.
However, spatial fluctuations in the temperature are natural if the stellar feedback scenario is correct. 
Indeed, the observed temperature range is consistent with the typical temperature range of  middle-aged Galactic supernova remnants.
Therefore, the high temperature regions might reflect fresh hot gases outflowing from the MW disk.
Another possibility is extragalactic hot gas associated with galaxy filaments \citep{2014ApJ...783..137M}.
Further observations covering large fractions of the blank X-ray sky are necessary to further examine the origins of these regions.

\section{Conclusions}
We derived the properties of the MW hot gaseous halo from an X-ray spectral analysis of 107 lines-of-sight from the Suzaku observations at $75\arcdeg<l<295\arcdeg$ and $|b|>15\arcdeg$.
The spectral model in the 0.4--5.0~keV band consists of three components: the hot gaseous halo component represented by a single-temperature CIE plasma, the local emission component empirically mimicked by a single-temperature CIE plasma, and the CXB component with a single power-law function.
We used the latest atomic database and solar abundance table, which affect the emission measure and the iron abundance.

The median temperature in the observed fields is 0.26~keV ($3.0\times10^6$~K), and the 16--84th percentile range is  0.19--0.32~keV (2.2--$3.8\times10^6$~K) showing a $\sim30$\% spatial fluctuation in the temperature.
The derived emission measure ranges over 0.6--$16.4\times10^{-3}$\EMunit{}.
We also constrained \Fehalo{} for the 46 lines-of-sight, and its median is 0.25.
The emission measure marginally correlates with $\csc |b|$.

The spatial distribution of \EMhalo{} is approximated by a disk-like density distribution with $n_0 \sim 4\times10^{-3}$~cm$^{-3}$, $R_0 \sim 7$~kpc, and $z_0 \sim 2$~kpc, even though there is a $\sim40$\% scatter of the data around the model.
We also found that the contribution of the extended spherical hot gas to the observed X-ray emission is minor but its mass contribution is much higher than that of the disk-like component.
This is because the X-ray flux, which is proportional to the square of the density, is biased toward high density regions.

The disk-like hot gas component likely results from stellar feedback in the MW disk, according to its small scale height and the large scatter of \EMhalo{}. 
The over-solar \Fehalo{} indicates a significant contribution of core-collapse supernovae and supports the stellar feedback scenario.

In addition, we found six lines-of-sight that has significantly high temperatures ($>0.4$~keV).
The possible origin of these high temperature regions is hot gas recently outflowing from the MW disk  and/or extragalactic hot gas filaments between galaxies.

\acknowledgments
We thank all of the Suzaku team members for developing hardwares and softwares, spacecraft operations, and instrument calibration.
We also thank T. Takahashi and K. Odaka for useful comments and discussion.
This work is supported by JSPS KAKENHI Grant Number 16K17674 (S.N.).

\software{Astropy \citep{2013A&A...558A..33A}, APEC \citep{2012ApJ...756..128F}, CIAO \citep{2006SPIE.6270E..1VF}, corner.py \citep{2016JOSS....1...24F}, emcee \citep{2013PASP..125..306F}, HEAsoft \citep[v6.22;][]{1996ASPC..101...17A}, Matplotlib \citep{2007CSE.....9...90H}, TBabs \citep[v2.3;][]{2000ApJ...542..914W}, xisnxbgen \citep{2008PASJ...60S..11T}, xisrmfgen and xissimarfgen \citep{2007PASJ...59S.113I}}

\bibliographystyle{aasjournal}
\bibliography{bibdesk}

%
%



\end{document}

%% file: table_pub_table_obs_results.tex
\begin{longrotatetable}
\startlongtable
\begin{deluxetable*}{crcrrrrrrrrrrrrrrc}
\setlength{\tabcolsep}{1.0mm}
\tabletypesize{\scriptsize}
\tablecaption{Observations and fitting results. \label{tab:obsresults}}

\tablehead{
\colhead{Region} & \colhead{Sequence} & \colhead{Target Name} 
& \colhead{$l$} & \colhead{$b$} & \colhead{$t$} & \colhead{DYE} & \colhead{$N_\mathrm{H}$}
& \colhead{$kT_\mathrm{halo}$} & \colhead{$EM_\mathrm{halo}$} & \colhead{[O/Fe]$_\mathrm{halo}$} & \colhead{$Z_\mathrm{halo}$} & \colhead{$S_\mathrm{halo}$}
& \colhead{$kT_\mathrm{local}$} & \colhead{$EM_\mathrm{local}$} & \colhead{$Z_\mathrm{local}$} 
& \colhead{$N_\mathrm{CXB}$} 
& \colhead{$C$/dof} 
}
\colnumbers
\startdata
1 & 802083010 & { \tiny COMABKG } & 75.73   & $83.17$ & 21.7 & $>$$20$ & 1.0 & $0.19_{-0.01}^{+0.16}$ & $0.6_{-0.4}^{+0.4}$ & $<$$0.92$ & 1 & 0.7 & 0.1 & $25.3_{-4.3}^{+3.3}$ & 1 & $8.0_{-0.3}^{+0.4}$ & 2858.7/2518 \\ 
2 & 403008010 & { \tiny AM HERCULES BGD } & 77.40   & $20.28$ & 44.3 & $>$$20$ & 6.5 & $0.29_{-0.03}^{+0.03}$ & $3.1_{-0.4}^{+0.4}$ & $>$$-0.03$ & 1 & 2.4 & 0.1 & $18.2_{-2.0}^{+2.0}$ & 1 & $9.8_{-0.3}^{+0.3}$ & 2636.1/2518 \\ 
3 & 704008010 & { \tiny 1739+518 } & 79.53   & $31.85$ & 16.9 & $>$$20$ & 3.1 & $0.30_{-0.03}^{+0.03}$ & $4.7_{-0.6}^{+0.7}$ & $>$$0.01$ & 1 & 3.7 & 0.1 & $26.8_{-3.9}^{+4.0}$ & 1 & $10.0_{-0.5}^{+0.5}$ & 2819.8/2518 \\ 
4 & 406007010 & { \tiny 1FGL J2339.7-0531 } & 81.35   & $-62.47$ & 89.0 & $>$$20$ & 3.2 & $0.15_{-0.00}^{+0.02}$ & $7.8_{-2.7}^{+0.5}$ & $<$$-0.20$ & 1 & 3.9 & 0.1 & $6.4_{-2.4}^{+5.1}$ & 1 & $9.9_{-0.1}^{+0.2}$ & 2743.7/2518 \\ 
5 & 707009010 & { \tiny 2FGL J0022.2-1853 } & 82.15   & $-79.37$ & 32.6 & $>$$30$ & 2.1 & $0.30_{-0.07}^{+0.05}$ & $2.3_{-0.5}^{+0.5}$ & $>$$-1.00$ & 1 & 2.0 & 0.1 & $<$$29.4$ & 1 & $11.9_{-0.4}^{+0.4}$ & 2702.8/2518 \\ 
6 & 709004010 & { \tiny SWIFT J2248.8+1725 } & 85.73   & $-36.41$ & 13.1 & $>$$90$ & 7.7 & $0.22_{-0.03}^{+0.01}$ & $5.9_{-0.8}^{+3.0}$ & $-0.04_{-0.25}^{+0.58}$ & 1 & 4.9 & 0.1 & $<$$27.0$ & 1 & $10.4_{-0.2}^{+0.2}$ & 2658.3/2518 \\ 
7 & 502047010 & { \tiny LOW\_LATITUDE\_86-21 } & 86.00   & $-20.79$ & 81.6 & $>$$20$ & 7.9 & $0.28_{-0.03}^{+0.01}$ & $5.1_{-0.3}^{+0.3}$ & $0.40_{-0.07}^{+0.13}$ & 1 & 4.2 & 0.1 & $20.7_{-1.3}^{+1.3}$ & 1 & $9.7_{-0.2}^{+0.2}$ & 2537.7/2518 \\ 
8 & 704050010 & { \tiny SDSS J1352+4239 } & 88.11   & $70.10$ & 21.3 & $>$$20$ & 1.0 & $0.22_{-0.03}^{+0.05}$ & $2.3_{-0.7}^{+0.9}$ & $>$$-0.96$ & 1 & 1.7 & 0.1 & $<$$30.7$ & 1 & $10.3_{-0.4}^{+0.4}$ & 2961.5/2518 \\ 
9 & 501005010 & { \tiny DRACO HVC REGION B } & 90.08   & $42.68$ & 61.6 & $>$$20$ & 1.5 & $0.18_{-0.01}^{+0.02}$ & $3.3_{-0.8}^{+0.8}$ & $<$$0.31$ & 1 & 2.3 & 0.1 & $14.4_{-2.9}^{+3.0}$ & 1 & $10.8_{-0.2}^{+0.2}$ & 2738.3/2518 \\ 
10 & 708023010 & { \tiny MRK533 } & 90.63   & $-48.79$ & 45.9 & $>$$50$ & 5.2 & $0.31_{-0.02}^{+0.02}$ & $3.6_{-0.4}^{+0.4}$ & $0.37_{-0.10}^{+0.13}$ & 1 & 3.1 & 0.1 & $27.1_{-2.1}^{+2.1}$ & 1 & $10.6_{-0.3}^{+0.3}$ & 2809.9/2518 \\ 
11 & 501004010 & { \tiny DRACO HVC REGION A } & 91.21   & $42.38$ & 61.2 & $>$$20$ & 1.8 & $0.24_{-0.03}^{+0.03}$ & $1.8_{-0.3}^{+0.3}$ & $0.11_{-0.29}^{+0.60}$ & 1 & 1.5 & 0.1 & $21.1_{-1.8}^{+1.8}$ & 1 & $11.4_{-0.3}^{+0.3}$ & 2675.4/2518 \\ 
12 & 904001010 & { \tiny GRB 090709A } & 91.79   & $20.21$ & 37.1 & $>$$20$ & 8.5 & $0.31_{-0.03}^{+0.03}$ & $3.6_{-0.6}^{+0.6}$ & $>$$-0.12$ & 1 & 3.0 & 0.1 & $15.3_{-2.4}^{+2.4}$ & 1 & $12.4_{-0.4}^{+0.4}$ & 2877.9/2518 \\ 
13 & 707008010 & { \tiny 2FGL J1502.1+5548 } & 92.73   & $52.90$ & 23.5 & $>$$30$ & 1.4 & $0.14_{-0.00}^{+0.04}$ & $9.8_{-6.3}^{+2.4}$ & $<$$0.76$ & 1 & 4.1 & 0.1 & $<$$38.8$ & 1 & $10.1_{-0.3}^{+0.2}$ & 2737.5/2518 \\ 
14 & 501101010 & { \tiny DRACO ENHANCEMENT } & 93.99   & $43.99$ & 33.8 & $>$$20$ & 1.1 & $0.38_{-0.04}^{+0.04}$ & $2.8_{-0.4}^{+0.4}$ & $>$$0.22$ & 1 & 2.2 & 0.1 & $23.1_{-1.9}^{+2.0}$ & 1 & $9.6_{-0.4}^{+0.4}$ & 2661.1/2518 \\ 
15 & 708026010 & { \tiny NGC 235A } & 94.13   & $-85.92$ & 13.2 & $>$$60$ & 1.5 & $0.26_{-0.04}^{+0.07}$ & $1.5_{-0.6}^{+0.8}$ & 0.0 (fixed) & 1 & 1.4 & 0.1 & $24.9_{-4.6}^{+3.9}$ & 1 & $9.4_{-0.4}^{+0.4}$ & 2849.5/2519 \\ 
16 & 100018010 & { \tiny NEP } & 95.75   & $28.68$ & 88.4 & $>$$20$ & 4.0 & $0.20_{-0.00}^{+0.00}$ & $9.0_{-0.6}^{+0.6}$ & $0.26_{-0.10}^{+0.11}$ & 1 & 6.1 & 0.1 & $13.4_{-1.5}^{+1.6}$ & 1 & $11.9_{-0.2}^{+0.2}$ & 5540.1/5040 \\ 
$\cdots$ & 500026010 & { \tiny NEP } & 95.79   & $28.66$ & 26.4 & $>$$20$ & 4.0 & $\cdots$ & $\cdots$ & $\cdots$ & $\cdots$ & $\cdots$ & $\cdots$ & $\cdots$ & $\cdots$ & $\cdots$ & $\cdots$ \\ 
17 & 504070010 & { \tiny NEP \#1 } & 96.38   & $29.79$ & 50.0 & $>$$30$ & 4.5 & $0.27_{-0.02}^{+0.01}$ & $1.9_{-0.2}^{+0.3}$ & $0.25_{-0.12}^{+0.16}$ & 1 & 1.6 & 0.1 & $15.9_{-1.4}^{+1.0}$ & 1 & $8.5_{-0.2}^{+0.2}$ & 10781.8/10084 \\ 
$\cdots$ & 504072010 & { \tiny NEP \#2 } & 96.39   & $29.79$ & 47.7 & $>$$20$ & 4.5 & $\cdots$ & $\cdots$ & $\cdots$ & $\cdots$ & $\cdots$ & $\cdots$ & $\cdots$ & $\cdots$ & $\cdots$ & $\cdots$ \\ 
$\cdots$ & 504074010 & { \tiny NEP \#3 } & 96.39   & $29.79$ & 42.5 & $>$$20$ & 4.5 & $\cdots$ & $\cdots$ & $\cdots$ & $\cdots$ & $\cdots$ & $\cdots$ & $\cdots$ & $\cdots$ & $\cdots$ & $\cdots$ \\ 
$\cdots$ & 504076010 & { \tiny NEP \#4 } & 96.40   & $29.79$ & 49.8 & $>$$20$ & 4.5 & $\cdots$ & $\cdots$ & $\cdots$ & $\cdots$ & $\cdots$ & $\cdots$ & $\cdots$ & $\cdots$ & $\cdots$ & $\cdots$ \\ 
18 & 100030020 & { \tiny A2218\_offset } & 97.72   & $40.12$ & 46.2 & $>$$20$ & 2.4 & $0.24_{-0.03}^{+0.04}$ & $1.5_{-0.3}^{+0.3}$ & $>$$-0.45$ & 1 & 1.1 & 0.1 & $15.3_{-1.4}^{+1.4}$ & 1 & $10.5_{-0.3}^{+0.3}$ & 2552.1/2518 \\ 
19 & 408030010 & { \tiny SWIFT J2319.4+2619 } & 98.48   & $-32.22$ & 20.0 & $>$$60$ & 6.8 & $0.16_{-0.00}^{+0.02}$ & $12.5_{-2.6}^{+3.2}$ & $<$$-0.15$ & 1 & 6.6 & 0.1 & $<$$36.6$ & 1 & $10.8_{-0.4}^{+0.4}$ & 2848.0/2518 \\ 
20 & 705027010 & { \tiny EMS1341 } & 102.86   & $19.44$ & 14.1 & $>$$20$ & 21.0 & $0.31_{-0.03}^{+0.04}$ & $5.6_{-1.2}^{+1.2}$ & $>$$-0.32$ & 1 & 4.9 & 0.1 & $<$$13.9$ & 1 & $10.6_{-0.6}^{+0.6}$ & 2778.4/2518 \\ 
21 & 704014010 & { \tiny UGC 12741 } & 105.66   & $-29.88$ & 48.0 & $>$$20$ & 7.9 & $0.24_{-0.02}^{+0.03}$ & $3.2_{-0.6}^{+0.6}$ & $<$$0.28$ & 1 & 3.1 & 0.1 & $<$$15.8$ & 1 & $9.8_{-0.3}^{+0.3}$ & 2838.8/2518 \\ 
22 & 705023010 & { \tiny LEDA 84274 } & 106.76   & $47.40$ & 49.5 & $>$$20$ & 1.3 & $0.71_{-0.06}^{+0.05}$ & $2.0_{-0.5}^{+0.5}$ & $0.36_{-0.14}^{+0.23}$ & 1 & 2.2 & 0.1 & $18.4_{-1.8}^{+1.8}$ & 1 & $9.6_{-0.3}^{+0.3}$ & 2758.0/2518 \\ 
23 & 403039010 & { \tiny ASAS J002511+1217.2 } & 112.92   & $-50.08$ & 33.2 & $>$$20$ & 5.7 & $0.26_{-0.04}^{+0.05}$ & $2.5_{-0.6}^{+0.7}$ & $>$$-0.60$ & 1 & 2.0 & 0.1 & $<$$18.0$ & 1 & $14.1_{-0.4}^{+0.4}$ & 2726.8/2518 \\ 
24 & 705046010 & { \tiny IRAS 00397-1312 } & 113.89   & $-75.66$ & 58.5 & $>$$20$ & 1.8 & $0.24_{-0.09}^{+0.06}$ & $1.7_{-0.5}^{+3.2}$ & $<$$0.97$ & 1 & 1.5 & 0.1 & $<$$20.0$ & 1 & $10.6_{-0.3}^{+0.3}$ & 2760.8/2518 \\ 
25 & 407039010 & { \tiny EUVE J1439 +75.0 } & 114.11   & $40.14$ & 10.9 & $>$$50$ & 3.3 & $0.25_{-0.03}^{+0.03}$ & $3.4_{-0.9}^{+1.1}$ & $0.05_{-0.28}^{+0.49}$ & 1 & 3.1 & 0.1 & $22.0_{-5.5}^{+5.3}$ & 1 & $7.8_{-0.3}^{+0.3}$ & 2865.2/2518 \\ 
26 & 706005010 & { \tiny NGC6251\_LOBE\_BGD2 } & 115.82   & $31.61$ & 10.8 & $>$$20$ & 6.0 & $0.31_{-0.03}^{+0.03}$ & $3.9_{-0.6}^{+0.6}$ & $0.22_{-0.16}^{+0.23}$ & 1 & 3.6 & 0.1 & $19.2_{-3.0}^{+3.1}$ & 1 & $9.3_{-0.5}^{+0.5}$ & 5618.3/5040 \\ 
$\cdots$ & 706005020 & { \tiny NGC6251\_LOBE\_BGD2 } & 115.79   & $31.61$ & 11.2 & $>$$20$ & 6.0 & $\cdots$ & $\cdots$ & $\cdots$ & $\cdots$ & $\cdots$ & $\cdots$ & $\cdots$ & $\cdots$ & $\cdots$ & $\cdots$ \\ 
27 & 706004010 & { \tiny NGC6251\_LOBE\_BGD1 } & 116.19   & $31.04$ & 18.8 & $>$$20$ & 7.9 & $0.29_{-0.03}^{+0.03}$ & $4.3_{-0.7}^{+0.7}$ & $>$$-0.35$ & 1 & 3.6 & 0.1 & $32.6_{-3.5}^{+3.6}$ & 1 & $8.4_{-0.4}^{+0.4}$ & 2924.4/2518 \\ 
28 & 705012010 & { \tiny EMS1160 } & 120.03   & $27.94$ & 18.2 & $>$$30$ & 8.6 & $0.32_{-0.05}^{+0.08}$ & $2.3_{-0.7}^{+0.8}$ & $<$$0.87$ & 1 & 2.4 & 0.1 & $24.6_{-3.5}^{+3.6}$ & 1 & $12.1_{-0.5}^{+0.5}$ & 2874.4/2518 \\ 
29 & 405034010 & { \tiny EG AND } & 121.55   & $-22.18$ & 100.3 & $>$$20$ & 13.0 & $0.29_{-0.01}^{+0.01}$ & $5.2_{-0.4}^{+0.4}$ & $0.23_{-0.09}^{+0.11}$ & 1 & 4.6 & 0.1 & $28.3_{-1.5}^{+1.5}$ & 1 & $9.3_{-0.2}^{+0.2}$ & 2603.7/2518 \\ 
30 & 706037010 & { \tiny MRK 231 } & 121.76   & $60.26$ & 83.9 & $>$$50$ & 1.0 & $0.25_{-0.05}^{+0.07}$ & $0.9_{-0.4}^{+0.4}$ & $<$$0.51$ & 1 & 0.9 & 0.1 & $24.7_{-2.4}^{+2.4}$ & 1 & $10.7_{-0.2}^{+0.2}$ & 2645.9/2518 \\ 
31 & 708039010 & { \tiny VII ZW 403 } & 127.83   & $37.32$ & 67.3 & $>$$50$ & 3.9 & $0.19_{-0.01}^{+0.01}$ & $7.3_{-1.4}^{+1.4}$ & $<$$0.11$ & 1 & 5.2 & 0.1 & $<$$15.8$ & 1 & $11.3_{-0.2}^{+0.2}$ & 2619.2/2518 \\ 
32 & 705024010 & { \tiny IRAS 01250+2832 } & 132.51   & $-33.40$ & 57.6 & $>$$20$ & 8.2 & $0.26_{-0.03}^{+0.04}$ & $3.1_{-0.6}^{+0.6}$ & $0.21_{-0.31}^{+0.28}$ & 1 & 2.7 & 0.1 & $14.8_{-2.3}^{+2.6}$ & 1 & $9.7_{-0.3}^{+0.3}$ & 2704.1/2518 \\ 
33 & 709003010 & { \tiny NGC 2655 } & 134.94   & $32.69$ & 25.1 & $>$$70$ & 2.4 & $0.31_{-0.02}^{+0.02}$ & $3.8_{-0.4}^{+0.4}$ & $0.32_{-0.09}^{+0.11}$ & 1 & 3.3 & 0.1 & $26.2_{-2.7}^{+2.7}$ & 1 & $10.0_{-0.2}^{+0.2}$ & 2565.1/2518 \\ 
34 & 705054010 & { \tiny NGC 3147 } & 136.30   & $39.48$ & 120.1 & $>$$20$ & 3.3 & $0.26_{-0.02}^{+0.03}$ & $2.6_{-0.4}^{+0.4}$ & $0.24_{-0.15}^{+0.20}$ & 1 & 2.2 & 0.1 & $21.8_{-2.1}^{+2.3}$ & 1 & $10.9_{-0.2}^{+0.2}$ & 2554.8/2518 \\ 
35 & 505044010 & { \tiny L139\_B-32 } & 138.76   & $-32.31$ & 83.8 & $>$$20$ & 6.9 & $0.25_{-0.02}^{+0.03}$ & $2.4_{-0.4}^{+0.4}$ & $0.12_{-0.23}^{+0.35}$ & 1 & 2.1 & 0.1 & $16.5_{-1.9}^{+1.9}$ & 1 & $7.8_{-0.2}^{+0.2}$ & 2638.8/2518 \\ 
36 & 506025010 & { \tiny 3C 59 VICINITY 2 } & 141.95   & $-31.19$ & 125.6 & $>$$50$ & 6.6 & $0.24_{-0.01}^{+0.02}$ & $2.9_{-0.4}^{+0.4}$ & $0.05_{-0.19}^{+0.27}$ & 1 & 2.5 & 0.1 & $16.2_{-1.5}^{+1.5}$ & 1 & $9.4_{-0.2}^{+0.2}$ & 2690.8/2518 \\ 
37 & 506024010 & { \tiny 3C 59 VICINITY 1 } & 142.14   & $-29.91$ & 41.9 & $>$$40$ & 7.2 & $0.19_{-0.01}^{+0.01}$ & $5.9_{-1.5}^{+1.5}$ & 0.0 (fixed) & 1 & 3.9 & 0.1 & $<$$18.1$ & 1 & $7.5_{-0.2}^{+0.3}$ & 2654.1/2519 \\ 
38 & 407043010 & { \tiny CH UMA } & 142.91   & $42.66$ & 45.2 & $>$$20$ & 4.7 & $0.15_{-0.00}^{+0.02}$ & $10.9_{-4.2}^{+1.4}$ & 0.0 (fixed) & 1 & 4.4 & 0.1 & $<$$28.4$ & 1 & $9.9_{-0.3}^{+0.3}$ & 2843.7/2519 \\ 
39 & 803041010 & { \tiny NGC1961BACKGROUND } & 145.25   & $18.81$ & 24.1 & $>$$20$ & 13.1 & $0.25_{-0.02}^{+0.04}$ & $3.7_{-0.9}^{+0.8}$ & $>$$-0.61$ & 1 & 3.1 & 0.1 & $<$$7.4$ & 1 & $8.2_{-0.4}^{+0.4}$ & 2830.7/2518 \\ 
40 & 705003010 & { \tiny 1150+497 } & 145.52   & $64.98$ & 105.6 & $>$$20$ & 2.2 & $0.15_{-0.00}^{+0.02}$ & $5.8_{-2.4}^{+0.7}$ & $<$$0.05$ & 1 & 3.0 & 0.1 & $<$$25.0$ & 1 & $11.6_{-0.2}^{+0.2}$ & 2583.6/2518 \\ 
41 & 704048010 & { \tiny NGC 3718 } & 146.88   & $60.21$ & 49.9 & $>$$30$ & 1.1 & $0.22_{-0.03}^{+0.01}$ & $1.9_{-0.4}^{+1.3}$ & 0.0 (fixed) & 1 & 1.5 & 0.1 & $<$$16.1$ & 1 & $11.0_{-0.3}^{+0.3}$ & 2615.6/2519 \\ 
42 & 100046010 & { \tiny LOCKMANHOLE } & 148.98   & $53.15$ & 66.4 & $>$$30$ & 0.6 & $0.26_{-0.02}^{+0.01}$ & $1.4_{-0.1}^{+0.2}$ & $0.08_{-0.15}^{+0.21}$ & 1 & 1.3 & 0.1 & $16.2_{-1.0}^{+0.9}$ & 1 & $11.3_{-0.2}^{+0.2}$ & 13261.7/12606 \\ 
$\cdots$ & 101002010 & { \tiny LOCKMAN HOLE } & 149.70   & $53.20$ & 39.7 & $>$$40$ & 0.6 & $\cdots$ & $\cdots$ & $\cdots$ & $\cdots$ & $\cdots$ & $\cdots$ & $\cdots$ & $\cdots$ & $\cdots$ & $\cdots$ \\ 
$\cdots$ & 102018010 & { \tiny LOCKMANHOLE } & 149.71   & $53.19$ & 90.3 & $>$$30$ & 0.6 & $\cdots$ & $\cdots$ & $\cdots$ & $\cdots$ & $\cdots$ & $\cdots$ & $\cdots$ & $\cdots$ & $\cdots$ & $\cdots$ \\ 
$\cdots$ & 103009010 & { \tiny LOCKMANHOLE } & 149.70   & $53.20$ & 71.7 & $>$$20$ & 0.6 & $\cdots$ & $\cdots$ & $\cdots$ & $\cdots$ & $\cdots$ & $\cdots$ & $\cdots$ & $\cdots$ & $\cdots$ & $\cdots$ \\ 
$\cdots$ & 104002010 & { \tiny LOCKMAN HOLE } & 149.70   & $53.20$ & 92.1 & $>$$20$ & 0.6 & $\cdots$ & $\cdots$ & $\cdots$ & $\cdots$ & $\cdots$ & $\cdots$ & $\cdots$ & $\cdots$ & $\cdots$ & $\cdots$ \\ 
43 & 504062010 & { \tiny VICINITY OF NGC 4051 } & 150.13   & $70.30$ & 89.5 & $>$$20$ & 1.2 & $0.29_{-0.03}^{+0.03}$ & $1.9_{-0.3}^{+0.3}$ & $0.33_{-0.15}^{+0.22}$ & 1 & 1.6 & 0.1 & $16.3_{-1.7}^{+1.7}$ & 1 & $10.8_{-0.2}^{+0.2}$ & 2509.1/2518 \\ 
44 & 704013010 & { \tiny 2MASX J02485937+2630 } & 153.13   & $-29.32$ & 43.2 & $>$$20$ & 15.2 & $0.30_{-0.04}^{+0.04}$ & $2.4_{-0.6}^{+0.6}$ & $0.02_{-0.30}^{+0.50}$ & 1 & 2.5 & 0.1 & $16.1_{-1.9}^{+1.9}$ & 1 & $10.5_{-0.3}^{+0.4}$ & 2714.9/2518 \\ 
45 & 705001010 & { \tiny MRK 18 } & 155.86   & $39.40$ & 38.0 & $>$$20$ & 5.0 & $0.25_{-0.05}^{+0.07}$ & $1.7_{-0.7}^{+1.6}$ & 0.0 (fixed) & 1 & 1.6 & 0.1 & $<$$28.3$ & 1 & $12.5_{-0.4}^{+0.4}$ & 2761.7/2519 \\ 
46 & 707021010 & { \tiny AO 0235+164 } & 156.78   & $-39.11$ & 26.2 & $>$$50$ & 10.3 & $0.18_{-0.02}^{+0.02}$ & $5.9_{-2.1}^{+3.6}$ & 0.0 (fixed) & 1 & 3.6 & 0.1 & $<$$21.6$ & 1 & $8.8_{-0.3}^{+0.3}$ & 2725.8/2519 \\ 
47 & 501104010 & { \tiny MBM12 OFF-CLOUD } & 157.34   & $-36.82$ & 20.1 & $>$$20$ & 9.0 & $0.20_{-0.02}^{+0.03}$ & $3.1_{-1.2}^{+1.5}$ & 0.0 (fixed) & 1 & 2.2 & 0.1 & $<$$17.6$ & 1 & $7.5_{-0.4}^{+0.4}$ & 2745.4/2519 \\ 
48 & 402046010 & { \tiny BZ UMA } & 159.02   & $38.83$ & 29.7 & $>$$20$ & 4.8 & $0.34_{-0.06}^{+0.06}$ & $2.1_{-0.6}^{+0.6}$ & $>$$-0.57$ & 1 & 1.6 & 0.1 & $11.6_{-2.7}^{+2.5}$ & 1 & $11.8_{-0.6}^{+0.6}$ & 2756.5/2518 \\ 
49 & 709021010 & { \tiny I ZW 18 } & 160.54   & $44.84$ & 16.6 & $>$$30$ & 2.7 & $0.14_{-0.00}^{+0.01}$ & $16.4_{-4.5}^{+1.9}$ & 0.0 (fixed) & 1 & 5.5 & 0.1 & $<$$29.6$ & 1 & $8.2_{-0.4}^{+0.3}$ & 2831.9/2519 \\ 
50 & 703065010 & { \tiny IRASF01475-0740 } & 160.70   & $-65.86$ & 57.9 & $>$$20$ & 2.2 & $0.22_{-0.03}^{+0.06}$ & $<$$4.7$ & 0.0 (fixed) & 1 & 0.5 & 0.1 & $<$$14.3$ & 1 & $9.3_{-0.3}^{+0.3}$ & 2738.0/2519 \\ 
51 & 704052010 & { \tiny SDSS J0943+5417 } & 161.23   & $46.42$ & 34.2 & $>$$20$ & 1.5 & $0.34_{-0.04}^{+0.05}$ & $2.0_{-0.4}^{+0.4}$ & $>$$-0.14$ & 1 & 1.7 & 0.1 & $16.1_{-2.5}^{+2.2}$ & 1 & $10.0_{-0.3}^{+0.4}$ & 2700.7/2518 \\ 
52 & 709019010 & { \tiny Q0142-100 } & 161.64   & $-68.48$ & 56.7 & $>$$30$ & 3.2 & $0.19_{-0.03}^{+0.10}$ & $2.8_{-1.6}^{+3.4}$ & $<$$0.62$ & 1 & 2.1 & 0.1 & $11.4_{-6.4}^{+4.8}$ & 1 & $9.7_{-0.2}^{+0.2}$ & 2623.4/2518 \\ 
53 & 402044010 & { \tiny SW UMA } & 164.81   & $36.96$ & 16.9 & $>$$20$ & 4.1 & $0.65_{-0.23}^{+0.09}$ & $3.4_{-0.8}^{+0.8}$ & $>$$-0.10$ & 1 & 2.9 & 0.1 & $14.0_{-2.4}^{+2.5}$ & 1 & $7.8_{-0.6}^{+0.6}$ & 2799.3/2518 \\ 
54 & 509008010 & { \tiny HOT BLOB 2 } & 164.90   & $38.21$ & 21.4 & $>$$80$ & 3.2 & $0.23_{-0.02}^{+0.02}$ & $3.4_{-0.7}^{+0.9}$ & $-0.21_{-0.44}^{+0.69}$ & 1 & 3.1 & 0.1 & $16.1_{-3.5}^{+3.3}$ & 1 & $9.5_{-0.2}^{+0.2}$ & 2598.3/2518 \\ 
55 & 701057010 & { \tiny APM 08279+5255 } & 165.74   & $36.24$ & 85.2 & $>$$20$ & 4.7 & $0.31_{-0.02}^{+0.02}$ & $1.7_{-0.1}^{+0.1}$ & $0.35_{-0.08}^{+0.10}$ & 1 & 1.5 & 0.1 & $12.9_{-0.7}^{+0.7}$ & 1 & $10.4_{-0.2}^{+0.2}$ & 7972.2/7562 \\ 
$\cdots$ & 701057020 & { \tiny APM 08279+5255 } & 165.74   & $36.24$ & 64.1 & $>$$30$ & 4.7 & $\cdots$ & $\cdots$ & $\cdots$ & $\cdots$ & $\cdots$ & $\cdots$ & $\cdots$ & $\cdots$ & $\cdots$ & $\cdots$ \\ 
$\cdots$ & 701057030 & { \tiny APM 08279+5255 } & 165.76   & $36.24$ & 104.3 & $>$$20$ & 4.7 & $\cdots$ & $\cdots$ & $\cdots$ & $\cdots$ & $\cdots$ & $\cdots$ & $\cdots$ & $\cdots$ & $\cdots$ & $\cdots$ \\ 
56 & 508073010 & { \tiny MBM16-OFF } & 165.86   & $-38.39$ & 78.2 & $>$$30$ & 19.0 & $0.19_{-0.02}^{+0.04}$ & $5.4_{-2.0}^{+2.2}$ & $<$$0.91$ & 1 & 3.7 & 0.1 & $9.7_{-2.1}^{+2.2}$ & 1 & $8.8_{-0.2}^{+0.2}$ & 2568.6/2518 \\ 
57 & 505058010 & { \tiny L168\_B53 } & 167.64   & $53.19$ & 46.3 & $>$$40$ & 0.9 & $0.18_{-0.01}^{+0.01}$ & $4.4_{-1.3}^{+1.3}$ & 0.0 (fixed) & 1 & 2.7 & 0.1 & $<$$26.9$ & 1 & $8.3_{-0.2}^{+0.2}$ & 2463.3/2519 \\ 
58 & 509009010 & { \tiny HOT BLOB 3 } & 167.88   & $36.01$ & 18.4 & $>$$80$ & 5.0 & $0.24_{-0.03}^{+0.04}$ & $1.8_{-0.7}^{+0.8}$ & $<$$0.36$ & 1 & 1.7 & 0.1 & $19.5_{-3.3}^{+3.3}$ & 1 & $8.0_{-0.2}^{+0.2}$ & 2585.9/2518 \\ 
59 & 703042010 & { \tiny J081618.99+482328.4 } & 171.02   & $33.70$ & 90.9 & $>$$20$ & 5.8 & $0.28_{-0.02}^{+0.03}$ & $1.6_{-0.3}^{+0.3}$ & $0.09_{-0.23}^{+0.31}$ & 1 & 1.6 & 0.1 & $8.9_{-0.6}^{+1.3}$ & 1 & $10.5_{-0.2}^{+0.2}$ & 2612.3/2518 \\ 
60 & 709009010 & { \tiny ARP318 } & 173.96   & $-64.97$ & 77.3 & $>$$30$ & 2.8 & $0.32_{-0.02}^{+0.02}$ & $3.4_{-0.3}^{+0.3}$ & $0.17_{-0.10}^{+0.13}$ & 1 & 3.3 & 0.1 & $15.0_{-1.5}^{+1.5}$ & 1 & $10.9_{-0.2}^{+0.2}$ & 2717.3/2518 \\ 
61 & 703008010 & { \tiny SWIFT J0911.2+4533 } & 174.71   & $43.11$ & 76.6 & $>$$20$ & 1.3 & $0.22_{-0.02}^{+0.02}$ & $1.8_{-0.3}^{+0.5}$ & $<$$0.79$ & 1 & 1.5 & 0.1 & $<$$15.0$ & 1 & $9.7_{-0.2}^{+0.2}$ & 2628.6/2518 \\ 
62 & 706013010 & { \tiny 3C78 } & 174.85   & $-44.51$ & 96.1 & $>$$20$ & 14.6 & $0.39_{-0.02}^{+0.02}$ & $6.4_{-0.4}^{+0.4}$ & $0.62_{-0.04}^{+0.04}$ & 1 & 5.3 & 0.1 & $11.0_{-1.1}^{+1.2}$ & 1 & $9.9_{-0.2}^{+0.2}$ & 2714.8/2518 \\ 
63 & 706038010 & { \tiny IRAS 09104+4109 } & 180.99   & $43.55$ & 72.7 & $>$$40$ & 1.5 & $0.30_{-0.04}^{+0.03}$ & $1.8_{-0.3}^{+0.3}$ & $>$$-0.74$ & 1 & 1.4 & 0.1 & $17.5_{-1.8}^{+1.8}$ & 1 & $10.9_{-0.2}^{+0.2}$ & 2614.7/2518 \\ 
64 & 709007010 & { \tiny SWIFT J0714.2+3518 } & 182.49   & $19.57$ & 47.9 & $>$$90$ & 6.7 & $0.22_{-0.02}^{+0.02}$ & $2.5_{-0.5}^{+0.6}$ & $<$$0.30$ & 1 & 2.3 & 0.1 & $13.8_{-1.8}^{+1.9}$ & 1 & $8.4_{-0.2}^{+0.2}$ & 2819.4/2518 \\ 
65 & 704053010 & { \tiny IC 2497 } & 190.27   & $48.82$ & 76.3 & $>$$20$ & 1.1 & $0.33_{-0.03}^{+0.04}$ & $1.8_{-0.3}^{+0.3}$ & $>$$-0.08$ & 1 & 1.5 & 0.1 & $17.6_{-1.8}^{+1.8}$ & 1 & $10.5_{-0.2}^{+0.3}$ & 2631.1/2518 \\ 
66 & 707006010 & { \tiny 3C 236 BACKGROUND } & 190.35   & $53.69$ & 25.7 & $>$$30$ & 1.0 & $0.18_{-0.01}^{+0.02}$ & $4.9_{-1.6}^{+1.6}$ & 0.0 (fixed) & 1 & 3.1 & 0.1 & $<$$30.8$ & 1 & $8.4_{-0.3}^{+0.3}$ & 2863.1/2519 \\ 
67 & 708038010 & { \tiny IRAS F11119+3257 } & 192.21   & $68.35$ & 142.9 & $>$$70$ & 2.2 & $0.19_{-0.01}^{+0.02}$ & $4.1_{-0.8}^{+0.8}$ & $-0.11_{-0.36}^{+0.63}$ & 1 & 2.9 & 0.1 & $10.2_{-2.6}^{+2.6}$ & 1 & $9.2_{-0.1}^{+0.1}$ & 2608.1/2518 \\ 
68 & 409029010 & { \tiny 1RXS J032540.0-08144 } & 192.87   & $-48.95$ & 36.4 & $>$$30$ & 5.9 & $0.23_{-0.01}^{+0.01}$ & $15.6_{-0.9}^{+0.9}$ & $0.51_{-0.08}^{+0.09}$ & 1 & 11.7 & 0.1 & $22.0_{-3.4}^{+3.5}$ & 1 & $10.7_{-0.3}^{+0.3}$ & 2818.1/2518 \\ 
69 & 700011010 & { \tiny SWIFT J0746.3+2548 } & 194.52   & $22.92$ & 100.1 & $>$$20$ & 5.1 & $0.26_{-0.02}^{+0.02}$ & $2.2_{-0.2}^{+0.3}$ & $-0.01_{-0.21}^{+0.31}$ & 1 & 2.1 & 0.1 & $13.4_{-1.1}^{+1.1}$ & 1 & $10.8_{-0.2}^{+0.2}$ & 2728.2/2518 \\ 
70 & 703003010 & { \tiny Q0827+243 } & 200.02   & $31.88$ & 48.2 & $>$$20$ & 3.3 & $0.46_{-0.09}^{+0.13}$ & $1.4_{-0.4}^{+0.4}$ & $<$$0.90$ & 1 & 1.5 & 0.1 & $10.3_{-1.9}^{+1.8}$ & 1 & $11.7_{-0.4}^{+0.4}$ & 2572.3/2518 \\ 
71 & 407045010 & { \tiny BF ERI } & 201.04   & $-31.30$ & 28.3 & $>$$20$ & 5.8 & $0.30_{-0.02}^{+0.02}$ & $7.3_{-0.7}^{+0.7}$ & $>$$0.31$ & 1 & 5.5 & 0.1 & $31.6_{-3.1}^{+3.2}$ & 1 & $11.5_{-0.4}^{+0.4}$ & 2876.9/2518 \\ 
72 & 404035010 & { \tiny HD72779 } & 205.51   & $31.34$ & 71.0 & $>$$20$ & 2.9 & $0.64_{-0.05}^{+0.05}$ & $2.3_{-0.4}^{+0.4}$ & $0.45_{-0.08}^{+0.10}$ & 1 & 2.3 & 0.1 & $17.4_{-1.5}^{+1.5}$ & 1 & $8.2_{-0.3}^{+0.3}$ & 2632.5/2518 \\ 
73 & 408029010 & { \tiny V1159 ORI } & 206.53   & $-19.94$ & 76.0 & $>$$90$ & 27.6 & $0.31_{-0.01}^{+0.01}$ & $10.5_{-0.6}^{+0.6}$ & $0.33_{-0.04}^{+0.05}$ & 1 & 9.2 & 0.1 & $21.3_{-1.2}^{+1.2}$ & 1 & $12.0_{-0.2}^{+0.2}$ & 2623.6/2518 \\ 
74 & 708044010 & { \tiny B2 1023+25 } & 207.06   & $57.61$ & 59.5 & $>$$30$ & 1.7 & $0.18_{-0.01}^{+0.01}$ & $5.9_{-1.2}^{+1.1}$ & $<$$0.25$ & 1 & 3.9 & 0.1 & $<$$24.4$ & 1 & $8.6_{-0.2}^{+0.2}$ & 2749.9/2518 \\ 
75 & 702062010 & { \tiny Q0450-1310 } & 211.75   & $-32.07$ & 15.5 & $>$$20$ & 10.3 & $0.40_{-0.05}^{+0.13}$ & $3.6_{-0.7}^{+0.7}$ & $>$$-0.13$ & 1 & 3.4 & 0.1 & $15.0_{-2.7}^{+2.8}$ & 1 & $11.4_{-0.6}^{+0.6}$ & 2831.6/2518 \\ 
76 & 809052010 & { \tiny OFF-FIELD1 } & 212.25   & $55.01$ & 37.9 & $>$$40$ & 2.1 & $0.33_{-0.03}^{+0.06}$ & $2.9_{-0.4}^{+0.4}$ & $0.26_{-0.13}^{+0.19}$ & 1 & 2.7 & 0.1 & $24.4_{-2.3}^{+2.8}$ & 1 & $10.5_{-0.3}^{+0.3}$ & 2694.7/2518 \\ 
77 & 702115010 & { \tiny IRAS 10565+2448 } & 212.34   & $64.73$ & 39.4 & $>$$20$ & 1.1 & $0.35_{-0.04}^{+0.04}$ & $1.8_{-0.3}^{+0.3}$ & $0.33_{-0.13}^{+0.21}$ & 1 & 1.7 & 0.1 & $18.1_{-2.1}^{+1.7}$ & 1 & $8.1_{-0.3}^{+0.3}$ & 2721.9/2518 \\ 
78 & 502076010 & { \tiny ERIDANUS HOLE } & 213.44   & $-39.09$ & 103.7 & $>$$20$ & 2.6 & $0.26_{-0.02}^{+0.03}$ & $1.6_{-0.2}^{+0.3}$ & $0.05_{-0.25}^{+0.36}$ & 1 & 1.5 & 0.1 & $14.4_{-1.5}^{+1.4}$ & 1 & $7.9_{-0.2}^{+0.2}$ & 2591.2/2518 \\ 
79 & 707007010 & { \tiny 2FGL J0923.5+1508 } & 215.97   & $40.48$ & 91.5 & $>$$30$ & 3.2 & $0.15_{-0.01}^{+0.03}$ & $6.8_{-3.2}^{+3.1}$ & $<$$0.35$ & 1 & 3.2 & 0.1 & $<$$19.5$ & 1 & $10.6_{-0.2}^{+0.2}$ & 2519.9/2518 \\ 
80 & 409030010 & { \tiny IW ERIDANI } & 216.44   & $-40.61$ & 28.9 & $>$$40$ & 2.8 & $0.15_{-0.01}^{+0.01}$ & $8.9_{-1.1}^{+1.1}$ & 0.0 (fixed) & 1 & 3.7 & 0.1 & $<$$10.7$ & 1 & $8.7_{-0.3}^{+0.3}$ & 2939.0/2519 \\ 
81 & 708002010 & { \tiny NGC 3997 } & 218.72   & $77.83$ & 80.8 & $>$$40$ & 1.7 & $0.19_{-0.01}^{+0.02}$ & $5.6_{-2.1}^{+1.1}$ & $0.06_{-0.30}^{+0.47}$ & 1 & 3.8 & 0.1 & $<$$21.9$ & 1 & $10.6_{-0.2}^{+0.2}$ & 2739.3/2518 \\ 
82 & 704039010 & { \tiny PKS 0326-288 } & 224.90   & $-55.40$ & 56.5 & $>$$20$ & 1.0 & $0.20_{-0.02}^{+0.03}$ & $2.0_{-0.7}^{+1.2}$ & 0.0 (fixed) & 1 & 1.4 & 0.1 & $<$$22.2$ & 1 & $8.5_{-0.2}^{+0.2}$ & 2676.8/2519 \\ 
83 & 702076010 & { \tiny SWIFT J0918.5+0425 } & 227.10   & $34.42$ & 52.8 & $>$$20$ & 3.8 & $0.29_{-0.03}^{+0.03}$ & $1.8_{-0.3}^{+0.3}$ & $>$$-0.47$ & 1 & 1.5 & 0.1 & $12.3_{-1.7}^{+1.6}$ & 1 & $9.8_{-0.3}^{+0.3}$ & 2689.3/2518 \\ 
84 & 702064010 & { \tiny Q1017+1055 } & 230.36   & $50.83$ & 18.0 & $>$$20$ & 3.4 & $0.21_{-0.06}^{+0.04}$ & $<$$327.0$ & 0.0 (fixed) & 1 & 0.7 & 0.1 & $<$$16.0$ & 1 & $10.6_{-0.4}^{+0.5}$ & 2880.2/2519 \\ 
85 & 901005010 & { \tiny GRB070328 } & 235.19   & $-44.99$ & 52.6 & $>$$20$ & 2.9 & $0.40_{-0.05}^{+0.05}$ & $1.6_{-0.3}^{+0.3}$ & $0.35_{-0.12}^{+0.19}$ & 1 & 1.6 & 0.1 & $9.7_{-1.3}^{+1.3}$ & 1 & $9.1_{-0.3}^{+0.3}$ & 2815.5/2518 \\ 
86 & 709020020 & { \tiny HE0512-3329 } & 236.64   & $-33.86$ & 16.1 & $>$$50$ & 2.6 & $0.31_{-0.03}^{+0.04}$ & $2.6_{-0.4}^{+0.4}$ & $0.26_{-0.14}^{+0.20}$ & 1 & 2.4 & 0.1 & $26.3_{-2.4}^{+2.4}$ & 1 & $9.2_{-0.4}^{+0.4}$ & 5729.4/5040 \\ 
$\cdots$ & 709020030 & { \tiny HE0512-3329 } & 236.62   & $-33.85$ & 13.6 & $>$$60$ & 2.6 & $\cdots$ & $\cdots$ & $\cdots$ & $\cdots$ & $\cdots$ & $\cdots$ & $\cdots$ & $\cdots$ & $\cdots$ & $\cdots$ \\ 
87 & 506056010 & { \tiny G236+38 OFF } & 237.07   & $41.12$ & 62.5 & $>$$20$ & 2.1 & $0.18_{-0.03}^{+0.01}$ & $5.7_{-1.4}^{+4.8}$ & $>$$-0.85$ & 1 & 3.5 & 0.1 & $<$$23.2$ & 1 & $7.9_{-0.2}^{+0.2}$ & 2695.3/2518 \\ 
88 & 702031010 & { \tiny MRK 1239 } & 239.27   & $38.22$ & 26.1 & $>$$20$ & 4.4 & $0.32_{-0.07}^{+0.15}$ & $1.0_{-0.4}^{+0.4}$ & $<$$0.76$ & 1 & 1.2 & 0.1 & $23.0_{-2.3}^{+2.5}$ & 1 & $9.2_{-0.4}^{+0.4}$ & 2785.3/2518 \\ 
89 & 503104010 & { \tiny ARC\_BACKGROUND } & 240.49   & $-66.02$ & 167.0 & $>$$20$ & 4.1 & $0.27_{-0.04}^{+0.01}$ & $1.0_{-0.2}^{+0.3}$ & $0.02_{-0.24}^{+0.56}$ & 1 & 1.0 & 0.1 & $12.1_{-1.3}^{+1.0}$ & 1 & $8.5_{-0.2}^{+0.2}$ & 2508.9/2518 \\ 
90 & 405014010 & { \tiny PSR J0614-33 } & 240.50   & $-21.83$ & 31.1 & $>$$20$ & 3.9 & $0.35_{-0.02}^{+0.03}$ & $4.4_{-0.5}^{+0.5}$ & $0.37_{-0.09}^{+0.11}$ & 1 & 3.9 & 0.1 & $18.4_{-2.4}^{+2.5}$ & 1 & $8.3_{-0.4}^{+0.4}$ & 2813.0/2518 \\ 
91 & 703036020 & { \tiny Q0551-3637 } & 242.37   & $-26.92$ & 21.6 & $>$$20$ & 3.6 & $0.31_{-0.04}^{+0.04}$ & $2.8_{-0.5}^{+0.5}$ & $>$$-0.11$ & 1 & 2.1 & 0.1 & $16.0_{-2.8}^{+2.8}$ & 1 & $10.2_{-0.4}^{+0.5}$ & 2839.2/2518 \\ 
92 & 703040010 & { \tiny Q0940-1050 } & 246.39   & $30.44$ & 32.4 & $>$$20$ & 4.6 & $0.40_{-0.04}^{+0.04}$ & $2.7_{-0.4}^{+0.3}$ & $>$$0.23$ & 1 & 1.9 & 0.1 & $18.4_{-2.1}^{+2.1}$ & 1 & $11.1_{-0.4}^{+0.4}$ & 2656.8/2518 \\ 
93 & 703062010 & { \tiny NGC 1448 } & 251.60   & $-51.38$ & 53.0 & $>$$20$ & 1.0 & $0.20_{-0.04}^{+0.06}$ & $1.9_{-0.7}^{+3.0}$ & $<$$0.62$ & 1 & 1.6 & 0.1 & $<$$24.8$ & 1 & $10.9_{-0.3}^{+0.3}$ & 2513.7/2518 \\ 
94 & 703016010 & { \tiny SWIFT J0134.1-3625 } & 261.71   & $-77.06$ & 33.0 & $>$$40$ & 2.1 & $0.46_{-0.05}^{+0.06}$ & $2.7_{-0.4}^{+0.4}$ & $0.54_{-0.07}^{+0.08}$ & 1 & 2.4 & 0.1 & $15.9_{-2.0}^{+2.0}$ & 1 & $10.5_{-0.3}^{+0.3}$ & 2650.4/2518 \\ 
95 & 707012010 & { \tiny NGC 3431 } & 266.04   & $37.10$ & 55.1 & $>$$60$ & 4.8 & $0.15_{-0.01}^{+0.01}$ & $11.9_{-4.4}^{+4.3}$ & 0.0 (fixed) & 1 & 4.7 & 0.1 & $<$$27.4$ & 1 & $13.1_{-0.3}^{+0.3}$ & 2690.2/2519 \\ 
96 & 808057010 & { \tiny BULLET-BKG } & 266.15   & $-20.78$ & 43.1 & $>$$40$ & 6.8 & $0.25_{-0.02}^{+0.02}$ & $4.3_{-0.6}^{+0.6}$ & $0.28_{-0.15}^{+0.21}$ & 1 & 3.5 & 0.1 & $21.1_{-2.5}^{+2.5}$ & 1 & $8.6_{-0.2}^{+0.3}$ & 2753.6/2518 \\ 
97 & 708004010 & { \tiny ESO 119-G008 } & 266.67   & $-38.88$ & 44.6 & $>$$50$ & 1.3 & $0.23_{-0.02}^{+0.01}$ & $3.9_{-0.5}^{+1.2}$ & $0.08_{-0.20}^{+0.39}$ & 1 & 3.2 & 0.1 & $24.5_{-2.9}^{+2.9}$ & 1 & $8.2_{-0.2}^{+0.2}$ & 2634.3/2518 \\ 
98 & 708043010 & { \tiny NGC 3660 } & 269.10   & $48.36$ & 81.4 & $>$$50$ & 4.0 & $0.31_{-0.02}^{+0.02}$ & $4.0_{-0.3}^{+0.3}$ & $0.44_{-0.07}^{+0.08}$ & 1 & 3.3 & 0.1 & $24.1_{-1.7}^{+1.7}$ & 1 & $10.4_{-0.2}^{+0.2}$ & 2581.7/2518 \\ 
99 & 500027020 & { \tiny HIGH LAT. DIFFUSE B } & 272.40   & $-58.27$ & 50.7 & $>$$30$ & 3.3 & $0.33_{-0.03}^{+0.03}$ & $1.4_{-0.2}^{+0.2}$ & $>$$0.07$ & 1 & 1.1 & 0.1 & $12.3_{-1.1}^{+1.1}$ & 1 & $7.8_{-0.3}^{+0.3}$ & 2676.6/2518 \\ 
100 & 701008010 & { \tiny IRASF11223-1244 } & 272.55   & $44.74$ & 40.9 & $>$$20$ & 4.8 & $0.25_{-0.03}^{+0.04}$ & $2.0_{-0.5}^{+0.5}$ & $0.03_{-0.36}^{+0.73}$ & 1 & 1.8 & 0.1 & $20.3_{-2.2}^{+2.3}$ & 1 & $9.1_{-0.3}^{+0.3}$ & 2817.0/2518 \\ 
101 & 703037010 & { \tiny Q0109-3518 } & 275.46   & $-80.96$ & 30.0 & $>$$20$ & 2.0 & $0.25_{-0.06}^{+0.06}$ & $1.4_{-0.5}^{+0.9}$ & $-0.15_{-0.71}^{+2.68}$ & 1 & 1.4 & 0.1 & $19.5_{-3.8}^{+3.0}$ & 1 & $9.0_{-0.4}^{+0.4}$ & 2772.5/2518 \\ 
102 & 703002010 & { \tiny PKS0208-512 } & 276.10   & $-61.79$ & 51.9 & $>$$20$ & 1.9 & $0.22_{-0.02}^{+0.01}$ & $2.7_{-0.4}^{+1.1}$ & $>$$-0.82$ & 1 & 2.1 & 0.1 & $<$$19.4$ & 1 & $9.6_{-0.3}^{+0.3}$ & 2717.3/2518 \\ 
103 & 504069010 & { \tiny SEP \#1 } & 276.40   & $-29.82$ & 37.4 & $>$$20$ & 5.8 & $0.26_{-0.01}^{+0.02}$ & $5.3_{-0.4}^{+0.4}$ & $0.31_{-0.06}^{+0.07}$ & 1 & 4.4 & 0.1 & $22.2_{-1.7}^{+1.7}$ & 1 & $8.6_{-0.2}^{+0.2}$ & 11009.3/10084 \\ 
$\cdots$ & 504071010 & { \tiny SEP \#2 } & 276.40   & $-29.82$ & 52.5 & $>$$20$ & 5.8 & $\cdots$ & $\cdots$ & $\cdots$ & $\cdots$ & $\cdots$ & $\cdots$ & $\cdots$ & $\cdots$ & $\cdots$ & $\cdots$ \\ 
$\cdots$ & 504073010 & { \tiny SEP \#3 } & 276.39   & $-29.83$ & 40.8 & $>$$30$ & 5.8 & $\cdots$ & $\cdots$ & $\cdots$ & $\cdots$ & $\cdots$ & $\cdots$ & $\cdots$ & $\cdots$ & $\cdots$ & $\cdots$ \\ 
$\cdots$ & 504075010 & { \tiny SEP \#4 } & 276.39   & $-29.82$ & 49.7 & $>$$20$ & 5.8 & $\cdots$ & $\cdots$ & $\cdots$ & $\cdots$ & $\cdots$ & $\cdots$ & $\cdots$ & $\cdots$ & $\cdots$ & $\cdots$ \\ 
104 & 501002010 & { \tiny SKY\_53.3\_-63.4 } & 278.62   & $-45.31$ & 92.3 & $>$$20$ & 5.8 & $0.29_{-0.01}^{+0.01}$ & $4.5_{-0.2}^{+0.2}$ & $0.38_{-0.06}^{+0.07}$ & 1 & 3.7 & 0.1 & $16.1_{-1.0}^{+1.0}$ & 1 & $9.1_{-0.2}^{+0.2}$ & 2675.4/2518 \\ 
105 & 501001010 & { \tiny SKY\_50.0\_-62.4 } & 278.68   & $-47.08$ & 80.1 & $>$$20$ & 2.4 & $0.25_{-0.01}^{+0.01}$ & $5.2_{-0.3}^{+0.3}$ & $0.30_{-0.07}^{+0.09}$ & 1 & 4.2 & 0.1 & $27.9_{-1.7}^{+1.7}$ & 1 & $7.5_{-0.2}^{+0.2}$ & 2767.8/2518 \\ 
106 & 402089020 & { \tiny TW HYA } & 278.68   & $22.95$ & 20.0 & $>$$20$ & 6.8 & $0.28_{-0.02}^{+0.02}$ & $9.7_{-0.8}^{+0.9}$ & $0.28_{-0.11}^{+0.13}$ & 1 & 8.3 & 0.1 & $26.2_{-3.5}^{+3.6}$ & 1 & $10.9_{-0.5}^{+0.5}$ & 2875.2/2518 \\ 
107 & 705045010 & { \tiny IRAS 12072-0444 } & 283.97   & $56.32$ & 57.5 & $>$$20$ & 3.5 & $0.15_{-0.01}^{+0.02}$ & $8.4_{-3.7}^{+3.6}$ & $<$$0.15$ & 1 & 4.2 & 0.1 & $<$$28.1$ & 1 & $9.0_{-0.2}^{+0.2}$ & 2642.8/2518 \\ 
\enddata

\tablecomments{
(1) Region numbers. 
(2) Sequence numbers of the Suzaku archive.
(3) Target names shown in the event headers.
(4) Galactic longitude for the aim points in the unit of degree.
(5) Galactic latittude for the aim points in the unit of degree.
(6) Effective exposure times after the screening in the unit of ks.
(7) Screening criteria for the DYE\_ELV cut in the unit of degree.
(8) Fixed hydrogen column densities calculated according to \cite{2013MNRAS.431..394W} in the unit of $10^{20}$~cm$^{-2}$.
(9) Temperatures for the hot gaseous halo in the unit of keV.
(10) Emission measures for the hot gaseous halo in the unit of 10$^{-2}$~cm$^{-6}$~pc.
(11) Abundance ratios of oxygen to iron for the hot gaseous halo in the unit of dex.
(12) Metal abundance relative to the solar value execept for Fe.
(13) Unabsorbed surface brightness of the Galactic gasous halo component in the 0.4--1.0~keV band. The unit is $10^{-12}$~erg~cm$^{-2}$~s$^{-1}$~deg$^{2}$.
(14) Fixed temperatures for the local component in the unit of keV.
(15) Emission measures for the local component in the unit of 10$^{-2}$~cm$^{-6}$~pc.
(16) Fixed metal abundance relative to the solar value.
(17) Normalizations at 1~keV for the CXB component in the unit of ph~cm$^{-2}$~s$^{-1}$~sr$^{-1}$.
(18) Best-fit C-statistics and degree of freedom.
}
\end{deluxetable*}
\end{longrotatetable}